\documentstyle[12pt,epsf,axodraw]{article}
\begin{document}
\begin{center}
{\Large {\bf Code generation
(automatized programming) of symbolic formulae for helicity amplitudes}}

\vspace{0.5cm}

P.~Cherzor\footnote[1]{e-mail: cherzor@theory.sinp.msu.ru, cherzor@nikhef.nl}\\
\vspace{0.5cm}
{\it Skobeltsyn Institute of Nuclear Physics, Moscow State University \\
Moscow 119899, Russia }
\end{center}

\vspace{0.5cm}
\begin{abstract}
We propose a project for automatic representation and evaluation 
of helicity amplitudes we started to develop and explain it's 
main functioning principles.
\end{abstract}

\vspace{1cm} 

\noindent {\large {\bf Introduction}}

In the field of modern High
Energy Computational Physics (well-known computational packages
Comphep \cite{C}, Grace \cite{G}, FeynArts/FeynCalc\cite{F}, MADGRAPH \cite{MAD},
O'Mega \cite{omega}) a great attention is payed to the evaluation
of squared matrix elements of Feynman diagrams as a contribution
to the study of cross-section type observables.

In \cite{art1} we presented a
newly modified method for symbolic calculation of Feynman amplitudes. 
The idea of the method is to represent squared matrix elements through 
{\sl helicity amplitudes}.
The initial idea of the method rises to early 80th see
\cite{Kleiss85}, \cite{CALCUL81}. 
We choose a way to represent these helicity amplitudes in
a form, that can be used as a computer input and in the same way it is readable 
(comprehensible) by humans. As a programming tool we did use an
object-oriented programming language standard C++ \cite{book1}.
The idea of using helicity amplitudes and C++ language for generation
matrix elements for study multi-paricle production in high energy particles collisions
independently applied by authors of \cite{am++}.

There are three main aims of algorithmization: 
1) acceleration rapidity of calculations;
2) reducing an amount of auxiliary information, which requires to obtain results,
but does not contribute to them itself;
3) algorithmization in accordance with applicability for investigation concrete 
observables and possibility to determine them in an experiment.

We did make a program code to obtain symbolic formulae for helicity amplitudes,
that uses as an input data source the Comphep data for Feynman diagrams and 
produces as an output REDUCE program that contains helicity amplitudes in a form
of mathematical traces --- so-called {\sl fermion strings} in High Energy
Physics terminology \cite{art2}.

The essence of the algorithm to built helicity amplitudes expressions is the
following. For each diagram and for each polarization we have to obtain an
expression that is a product of $n$ ($n \ge 1$) fermion strings. The notion of
fermion strings was defined in \cite{art2} (expr.(1)).
As described in \cite{art1} one can assign {\sl basis spinor} 
$\xi _+$ or $\xi _-$ to any fermion. Plus or minus will be standing 
depending on the polarization ($\lambda = \pm 1$) and the type of the spinor 
($u$, $v$, $u^c$, $v^c$, $\bar u$, $\bar v$, $\bar {u^c}$, $\bar {v^c}$). 
So we regard spinor pairs as entities to form a 
{\sl short} or {\sl long (double)} fermion strings.
Our algorithm of fermion strings formation
gives tree level helicity amplitudes each of those contain 
not more than two polarization vectors 
($\eta$ and $k$ or $\eta^*$ and $k$ in one amplitude, not $\eta $
and $\eta^*$ together). To achive this form for helcity amplitudes one do not
need to apply any symbolic formulae substitutions (cf. \cite{art2}),
but to use traces building technic.

On the contrary, using algorithm of \cite{art2} one makes 
symbolic transformations of helicity amplitudes expressions
and obtains polarization vector dependence proportional either to $\eta $
and $\eta^*$.

\noindent {\large {\bf The realization of the algorithm to build helicity
amplitudes}}

On the step of forming
fermion strings we restricted our algorithm by a number of 
external particles equal 6. This includes most interesting processes, 
that can be evaluated, and fit for practical purposes. Further generalization
is straightforward, but analysis of results of output will be more 
cumbersome because of big number of amplitudes and a length of 
symbolic expressions, that will expire a lot of computer memory, 
and can be done later.
The algorithm to form fermion strings under above restrictions is the following:
 
\begin{itemize}
\item[1] First, one searches for a spinor pairs that can form short
fermion string, i.e., string, that contain only one pair of spinors 
(fermion and anti-fermion) of the same chirality.
\item[2] Then one searches if there exists a possibility to form a 
double (long)
fermion string, i.e., trace that contain four spinors (two 
spinor pairs each of them consists of fermion and anti-fermion).
\item [3] Finally, one should check if there any spinor pairs that need to be 
united into a fermion string yet, if needed short fermion strings arise.
\end{itemize}

What are the criteria to unify a pair of spinors into short or long fermion
string? We define fermion string as double if it contains two spinor pairs
so, that for each basis spinor in one pair it corresponds the basis spinor
of opposite chirality in the other pair. In each spinor pair that comprise
double fermion string the chiralities of basis spinors are opposite.

After defining all fermion strings in the amplitude a task reduces itself to
automatize a process of writing down fermion strings in a form of traces and 
further their multiplication to obtain amplitudes themselves.

But, what is the inner structure of a fermion string and how one can 
find it from Feynman diagram? How one can establish a squence of fermion
strings? 

The last question has a simple answer: sequence of fermion strings does not
strictly fixed, at least, if we have $g_{\mu \nu}=g_{\nu \mu}$ 
symmetric metric tensor in plain Minkowski space, 
i.e., changing an order of fermion strings does not change result
of evaluation, as each fermion string is a trace. Trace is a scalar value, 
or it does contain Lorentz indices, that can be contracted 
with ones from the other fermion string in the amplitude (more precisely, 
with metric tensors of
the numerator of boson particles propagators for Feynman gauge). All indices 
should be contracted inside helicity amplitude. We fix a sequence of fermion 
strings from the algorithm.

The algorithm of formation inner structure of
fermion string is done without strong limitation on the number of external
particles (it can be changed), 
structure of propagators and vertices (new structures can be added).
First we restricted vertex structure types by QED vertex structure and propagator
structure types by QED propagator in Feynman gauge for simplicity.
Let us consider electron-positron scattering $ee^+ \to ee^+\mu \mu ^+$ 
(see diagr.1 below). There are three spinor pairs in each whole helicity 
amplitude that related to this scattering process. 
Certainly, for polarized particles a fermion string structure of 
amplitude depends on a set of polarizations of external particles. E.g., if
one puts all polarizations equal $-1/2$ one will have three short fermion
strings in this helicity amplitude:
$$
A(\lambda_i=-1/2)\propto A_1 \,*\, A_2 \,*\, A_3, \;\;\; i=1,...6,\;\;\;
A_1=\bar {v}_{e^+}(p_2)\,\gamma_{\mu_1}\,v_{e^+}(p_4),
$$
$$
A_2=\bar {u}_{\mu }(p_5)\,\gamma_{\mu_2}\,v_{\mu ^+}(p_6), \;\;\;
A_3=\bar {u}_{e}(p_3)\,\gamma_{\mu_2}  \,prop({p}_8)\, \gamma_{\mu _1}\,u_{e}(p_1),
$$
where $p_8$ momentum of virtual electron, $prop(p_8)$ propagator of the virtual 
electron with momentum $p_8$. In our considerations we take into account only
numerator of particles propagators as denominators could be included to the
numerical factor, which is common for each helicity amplitude in the diagram.
One can found from the results of symbolic evaluation 
that any helicity amplitude of diagr.1 that contain long fermion
string nullify itself. It can be easily shown using trace mathematics
\cite{book2}, it occurs due to the {\sl chiral} symmetry 
of the part of each spinor product that depends on polarization. 
In this diagram only helicity amplitudes containing short fermion strings will
contribute. Moreover, amplitude constructions those contain more than one
$\eta $ ($\eta ^*$) contribution in one term also nullify itself.
It is a significant simplification one can observe.

For the Standard Model our code can process diagrams
with fermion and anti-fermion external particles, virtual photons and
bosons in Feynman gauge (excluding diagrams with Higgs
excluding also diagrams with many-gluon vertices 
if the QCD-sector of the Standard Model is involved).
Recently we did work on diagrams with external photons
(bosons) and planed to include also scalar particles, such as Higgs.
 
Comparing results obtained through summation over
all helicities for squared helicity amplitudes and non-polarized squared matrix 
elements for each diagram in a given
process (excluding cross terms), we did check numerically correctness of evaluation of 
helicity amplitudes using the algorithm to form fermion strings
on an examples of processes those comprise of tree level diagrams. 

\begin{itemize}
\item[1] First, for $2 \to 2$ scattering for the Standard Model process $ee^+ \to ee^+$ 
in the limit of massless electrons (positrons). No contradictions are found.
\item[2] In \cite{art2} we did describe processes $ee^+ \to t{\bar t}$ in the Standard Model
and $ee^+ \to neutralino \; pair$ in SUSY (MSSM) with massive quarks in final states,
the second process with heavy neutralinos, 
and massless limit for electron and positron mass(es) and tried
investigation of symbolic structure of helicity amplitudes  
on these simple examples. Numerical calculations give no contradictions.
\item[3] We tried the Standard Model diagrams (with $\gamma _\mu$ vertices) 
$ee^+ \to ee^+ \mu \mu^+$ process (assigning mass to muons and excluding 
diagrams with Higgs particles) (see diagr.1). The numerical results of 
summation over polarized
squared helicity amplitudes coincide with ones obtained for non-polarized
squared matrix elements, but one can put a question on the degree of confidence 
to machine calculated output. It is still remains difficulty to check a
correctness of symbolic formulae and numerical calculation for such a big
symbolic output.  
\item[4] We checked $e\gamma \to e\gamma $ (Compton scattering) 
very preliminarily, i.e., we did check results 
numerically in the limit of massless electrons (positrons).
\end{itemize}

Namely, we did treat numerically 
Lorentz covariant symbolic expressions for helicity amplitudes, 
obtained after evaluation traces 
(output of REDUCE code) of symbolic code for helicity amplitudes  
generated from our initial C++ program. 
We tend to call this our new computational project as CG-project 
(CG means {\sl CODE GENERATION}).
Briefly, one can characterize CG-project as a sequence of steps:
\begin{itemize}
\item[1] code on C++. Input: Comphep Feynman diagrams data. Output: REDUCE
program (formulae for helicity amplitudes).
\item[2] REDUCE program. Input: trace structures. Output: Lorentz covariant
symbolic expressions.
\end{itemize}

\noindent {\large {\bf CG-project realization}}

The realized part of work on creation 
CG-programming code could be formulated as an annex list of accomplished tasks.

\noindent 1) Using REDUCE package a set of test programs was created and an 
applicability of helicity amplitudes method 
for the evaluation of helicity amplitudes and squared matrix elements on the examples 
of smplest tree-level prosess $2 \to 2$ was tested (modification of helicity basis):

	a) numerical testing is applyed;

	b) symbolic testing is applyed;

	c) results' analysis is applyed.

\noindent 2) A way to optimize helicity amplitudes evaluation was found.

	a) It was created a set of programs allowing to test helicity amplitudes 
	optimization process and to compare
	results for helicity amplitudes in either numeric and symbolic form.

	b) Possibilities to optimize symbolic formulae for helicity amplitudes 
	were investigated on the examples
	of chosen diagrams of QED, SUSY (MSSM), Standard Model. 
	It was foreseeing a posibility to use C-conjugate 
	spinors. A posibilities to optimize symbolic formulae using C-conjugation
	and changing spinor pairs' conjunctions were taken into account.
	
	c) Analysis of the results of optimization of symbolic expressions for 
	helicity amplitudes was realized.

	d) The comparison for non-polarized squared matrix elements
	with Comphep package was realized in numeric and symbolic form.

\noindent 3) There were investigated ways to automatize helicity amplitudes evaluation for the 
case of multiple diagrams and
processes that includes $\le 6$ particles:

	a) initial code realized on programming language Standard C++ was created 
	with minimal user interface.
	A testing is applyed, an analysis of possibilities of optimization of the code 
	is applyed.

	b) Using object-oriented C++ features it was created a code that operates 
	with minimal test database 
	(that is a set of objects of a definite class) and generates file of 
	symbolic results (in the form of traces
	that one able to evaluate using REDUCE) for helicity amplitudes. 
	It was realized an analysis of the results
	of code functioning.

	c) The complarison of numerical values of non-polarized squared matrix elements, 
	those obtained by sum over polarizations of squared numerical 
	values for helicity amplitudes, with numerical values of non-polarized squared 
	matrix elements formulae was executed.
	
	d) The code for generation tree-level processes (which include up 
	to 6 particles in the initial and final states
	and have vertices of the type boson-fermion-antifermion) helicity 
	amplitudes' symbolic formulae was created. Applyed an algorithm of 
	optimization by changing spinor pair conjunction.

	e) An interface of our program, that generates trace symbolic 
	formulae for helicity
	amplitudes, with Comphep symbolic tables allowing to do 
	automatical generation of symbolic formulae for
	helicity amplitudes for mentioned above types of scattering processes' diagrams.

	f) It was applyed the testing on functioning of the code 
	and the interface with Comphep on the examples of selected
	process' diagrams and the comparison of numerical values 
	of non-polarized squared matrix elements of selected diagrams
	with results of summed over polarizations squared numerical values of
	symbolic formulae for helicity amplitudes. 
	
	g) It was created and tryed out the test program's interface: 
	debugger that functions on the first step of CG (see above)
	and translator for representation 
	symbolic formulae for helicity amplitudes in the form of traces 
	into LaTeX format (see Appendix: symbolic
	representation of helicity amplitudes for selected diagram of 
	the process $ee^+\to ee^+\mu {\mu }^+$). Translator can function
	when the first CG step is completed.\\	

\noindent {\large {\bf Conclusions}}

We described an algorithm to build helicity
amplitudes and a way we started to realize it.

It is not an accidental choice that we use C++ and REDUCE languages in our project.
We built a code that is, on one hand, in close connection with Comphep project, on the
other hand, C++ is the modest object-oriented language allowing more advanced interfaces
and structuring.

Our project on helicity amplitudes is at the begining. It could be expanded to
wider types of processes. It could be discussed a possibility of aplication it's
results to numerical computation of phenomenology of some High Energy or DIS processes.

\noindent {\large {\bf Acknowledgments}}

This work is supported by grants: RFFI 01-02-16710, INTAS CERN99-0377.

Sincere thanks to prof. A.~Pukhov for fruitful discussions and important
support and to prof. V.~Ilyin and prof. G. van der Steenhoven for valuable support
and possibility to make a part of work at NIKHEF.

\nocite{*} \bibliographystyle{aipproc}

\noindent
{\large {\bf Appendix}}

We give symbolic formulae for helicity amplitudes $A$
for diagram 1 (see an  applyed list of diagrams generated by Comphep) for the process
$ee^+\to ee^+\mu {\mu }^+$.
Here $m_0$ is the electron and positron mass, $q$ symbolizes the virtual electron momentum,
$M_m$ -- the mass of muon and corresponding anti-particle. One can find
$A$ dependence of initial and final states particles polarizations in round brackets
(positive or negative signs of polarizations $\lambda _i$):
$A(\lambda _1,\,\lambda _2,\,\lambda _3,\,\lambda _4,\,\lambda _5,\,\lambda _6)$.


$$A(-,-,-,-,-,-)=Tr\,[\hat {\mu _1}(-\hat {p_4}+m_0)(1-\gamma _5)\hat {k}(-\hat {p_2}+m_0)]$$
$$Tr\,[\hat {\mu _2}(-\hat {q}+m_0)\hat {\mu _1}(\hat {p_1}+m_0)(1+\gamma _5)\hat {k}(\hat {p_3}+m_0)]$$
$$Tr\,[\hat {\mu _2}(-\hat {p_6}+M_m)(1-\gamma _5)\hat {\eta ^*}\hat {k}(\hat {p_5}+M_m)];$$
$$A(-,-,-,-,-,+)=Tr\,[\hat {\mu _1}(-\hat {p_4}+m_0)(1-\gamma _5)\hat {k}(-\hat {p_2}+m_0)]$$
$$Tr\,[\hat {\mu _2}(-\hat {q}+m_0)\hat {\mu _1}(\hat {p_1}+m_0)(1+\gamma _5)\hat {k}(\hat {p_3}+m_0)]$$
$$Tr\,[\hat {\mu _2}(-\hat {p_6}+M_m)(1+\gamma _5)\hat {k}(\hat {p_5}+M_m)];$$
$$A(-,-,-,-,+,-)=Tr\,[\hat {\mu _1}(-\hat {p_4}+m_0)(1-\gamma _5)\hat {k}(-\hat {p_2}+m_0)]$$
$$Tr\,[\hat {\mu _2}(-\hat {q}+m_0)\hat {\mu _1}(\hat {p_1}+m_0)(1+\gamma _5)\hat {k}(\hat {p_3}+m_0)]$$
$$Tr\,[\hat {\mu _2}(-\hat {p_6}+M_m)(1-\gamma _5)\hat {k}(\hat {p_5}+M_m)];$$
$$A(-,-,-,-,+,+)=Tr\,[\hat {\mu _1}(-\hat {p_4}+m_0)(1-\gamma _5)\hat {k}(-\hat {p_2}+m_0)]$$
$$Tr\,[\hat {\mu _2}(-\hat {q}+m_0)\hat {\mu _1}(\hat {p_1}+m_0)(1+\gamma _5)\hat {k}(\hat {p_3}+m_0)]$$
$$Tr\,[\hat {\mu _2}(-\hat {p_6}+M_m)(-1)(1+\gamma _5)\hat {\eta }\hat {k}(\hat {p_5}+M_m)];$$
$$A(-,-,-,+,-,-)=Tr\,[\hat {\mu _1}(-\hat {p_4}+m_0)(1-\gamma _5)\hat {k}(-\hat {p_2}+m_0)]$$
$$Tr\,[\hat {\mu _2}(-\hat {q}+m_0)\hat {\mu _1}(\hat {p_1}+m_0)(1-\gamma _5)\hat {\eta ^*}\hat {k}(\hat {p_3}+m_0)]$$
$$Tr\,[\hat {\mu _2}(-\hat {p_6}+M_m)(1-\gamma _5)\hat {\eta ^*}\hat {k}(\hat {p_5}+M_m)];$$
$$A(-,-,-,+,-,+)=Tr\,[\hat {\mu _1}(-\hat {p_4}+m_0)(1-\gamma _5)\hat {k}(-\hat {p_2}+m_0)]$$
$$Tr\,[\hat {\mu _2}(-\hat {q}+m_0)\hat {\mu _1}(\hat {p_1}+m_0)(1-\gamma _5)\hat {\eta ^*}\hat {k}(\hat {p_3}+m_0)]$$
$$Tr\,[\hat {\mu _2}(-\hat {p_6}+M_m)(1+\gamma _5)\hat {k}(\hat {p_5}+M_m)];$$
$$A(-,-,-,+,+,-)=Tr\,[\hat {\mu _1}(-\hat {p_4}+m_0)(1-\gamma _5)\hat {k}(-\hat {p_2}+m_0)]$$
$$Tr\,[\hat {\mu _2}(-\hat {q}+m_0)\hat {\mu _1}(\hat {p_1}+m_0)(1-\gamma _5)\hat {\eta ^*}\hat {k}(\hat {p_3}+m_0)]$$
$$Tr\,[\hat {\mu _2}(-\hat {p_6}+M_m)(1-\gamma _5)\hat {k}(\hat {p_5}+M_m)];$$
$$A(-,-,-,+,+,+)=Tr\,[\hat {\mu _1}(-\hat {p_4}+m_0)(1-\gamma _5)\hat {k}(-\hat {p_2}+m_0)]$$
$$Tr\,[\hat {\mu _2}(-\hat {q}+m_0)\hat {\mu _1}(\hat {p_1}+m_0)(1-\gamma _5)\hat {k}(\hat {p_5}+M_m)$$
$$\hat {\mu _2}(-\hat {p_6}+M_m)(1+\gamma _5)\hat {k}(\hat {p_3}+m_0)];$$
$$A(-,-,+,-,-,-)=Tr\,[\hat {\mu _1}(-\hat {p_4}+m_0)(1-\gamma _5)\hat {k}(-\hat {p_2}+m_0)]$$
$$Tr\,[\hat {\mu _2}(-\hat {q}+m_0)\hat {\mu _1}(\hat {p_1}+m_0)(1+\gamma _5)\hat {k}(\hat {p_5}+M_m)$$
$$\hat {\mu _2}(-\hat {p_6}+M_m)(1-\gamma _5)\hat {k}(\hat {p_3}+m_0)];$$
$$A(-,-,+,-,-,+)=Tr\,[\hat {\mu _1}(-\hat {p_4}+m_0)(1-\gamma _5)\hat {k}(-\hat {p_2}+m_0)]$$
$$Tr\,[\hat {\mu _2}(-\hat {q}+m_0)\hat {\mu _1}(\hat {p_1}+m_0)(-1)(1+\gamma _5)\hat {\eta }\hat {k}(\hat {p_3}+m_0)]$$
$$Tr\,[\hat {\mu _2}(-\hat {p_6}+M_m)(1+\gamma _5)\hat {k}(\hat {p_5}+M_m)];$$
$$A(-,-,+,-,+,-)=Tr\,[\hat {\mu _1}(-\hat {p_4}+m_0)(1-\gamma _5)\hat {k}(-\hat {p_2}+m_0)]$$
$$Tr\,[\hat {\mu _2}(-\hat {q}+m_0)\hat {\mu _1}(\hat {p_1}+m_0)(-1)(1+\gamma _5)\hat {\eta }\hat {k}(\hat {p_3}+m_0)]$$
$$Tr\,[\hat {\mu _2}(-\hat {p_6}+M_m)(1-\gamma _5)\hat {k}(\hat {p_5}+M_m)];$$
$$A(-,-,+,-,+,+)=Tr\,[\hat {\mu _1}(-\hat {p_4}+m_0)(1-\gamma _5)\hat {k}(-\hat {p_2}+m_0)]$$
$$Tr\,[\hat {\mu _2}(-\hat {q}+m_0)\hat {\mu _1}(\hat {p_1}+m_0)(-1)(1+\gamma _5)\hat {\eta }\hat {k}(\hat {p_3}+m_0)]$$
$$Tr\,[\hat {\mu _2}(-\hat {p_6}+M_m)(-1)(1+\gamma _5)\hat {\eta }\hat {k}(\hat {p_5}+M_m)];$$
$$A(-,-,+,+,-,-)=Tr\,[\hat {\mu _1}(-\hat {p_4}+m_0)(1-\gamma _5)\hat {k}(-\hat {p_2}+m_0)]$$
$$Tr\,[\hat {\mu _2}(-\hat {q}+m_0)\hat {\mu _1}(\hat {p_1}+m_0)(1-\gamma _5)\hat {k}(\hat {p_3}+m_0)]$$
$$Tr\,[\hat {\mu _2}(-\hat {p_6}+M_m)(1-\gamma _5)\hat {\eta ^*}\hat {k}(\hat {p_5}+M_m)];$$
$$A(-,-,+,+,-,+)=Tr\,[\hat {\mu _1}(-\hat {p_4}+m_0)(1-\gamma _5)\hat {k}(-\hat {p_2}+m_0)]$$
$$Tr\,[\hat {\mu _2}(-\hat {q}+m_0)\hat {\mu _1}(\hat {p_1}+m_0)(1-\gamma _5)\hat {k}(\hat {p_3}+m_0)]$$
$$Tr\,[\hat {\mu _2}(-\hat {p_6}+M_m)(1+\gamma _5)\hat {k}(\hat {p_5}+M_m)];$$
$$A(-,-,+,+,+,-)=Tr\,[\hat {\mu _1}(-\hat {p_4}+m_0)(1-\gamma _5)\hat {k}(-\hat {p_2}+m_0)]$$
$$Tr\,[\hat {\mu _2}(-\hat {q}+m_0)\hat {\mu _1}(\hat {p_1}+m_0)(1-\gamma _5)\hat {k}(\hat {p_3}+m_0)]$$
$$Tr\,[\hat {\mu _2}(-\hat {p_6}+M_m)(1-\gamma _5)\hat {k}(\hat {p_5}+M_m)];$$
$$A(-,-,+,+,+,+)=Tr\,[\hat {\mu _1}(-\hat {p_4}+m_0)(1-\gamma _5)\hat {k}(-\hat {p_2}+m_0)]$$
$$Tr\,[\hat {\mu _2}(-\hat {q}+m_0)\hat {\mu _1}(\hat {p_1}+m_0)(1-\gamma _5)\hat {k}(\hat {p_3}+m_0)]$$
$$Tr\,[\hat {\mu _2}(-\hat {p_6}+M_m)(-1)(1+\gamma _5)\hat {\eta }\hat {k}(\hat {p_5}+M_m)];$$
$$A(-,+,-,-,-,-)=Tr\,[\hat {\mu _1}(-\hat {p_4}+m_0)(1+\gamma _5)\hat {k}(\hat {p_5}+M_m)$$
$$\hat {\mu _2}(-\hat {p_6}+M_m)(1-\gamma _5)\hat {k}(-\hat {p_2}+m_0)]$$
$$Tr\,[\hat {\mu _2}(-\hat {q}+m_0)\hat {\mu _1}(\hat {p_1}+m_0)(1+\gamma _5)\hat {k}(\hat {p_3}+m_0)];$$
$$A(-,+,-,-,-,+)=Tr\,[\hat {\mu _1}(-\hat {p_4}+m_0)(-1)(1+\gamma _5)\hat {\eta }\hat {k}(-\hat {p_2}+m_0)]$$
$$Tr\,[\hat {\mu _2}(-\hat {q}+m_0)\hat {\mu _1}(\hat {p_1}+m_0)(1+\gamma _5)\hat {k}(\hat {p_3}+m_0)]$$
$$Tr\,[\hat {\mu _2}(-\hat {p_6}+M_m)(1+\gamma _5)\hat {k}(\hat {p_5}+M_m)];$$
$$A(-,+,-,-,+,-)=Tr\,[\hat {\mu _1}(-\hat {p_4}+m_0)(-1)(1+\gamma _5)\hat {\eta }\hat {k}(-\hat {p_2}+m_0)]$$
$$Tr\,[\hat {\mu _2}(-\hat {q}+m_0)\hat {\mu _1}(\hat {p_1}+m_0)(1+\gamma _5)\hat {k}(\hat {p_3}+m_0)]$$
$$Tr\,[\hat {\mu _2}(-\hat {p_6}+M_m)(1-\gamma _5)\hat {k}(\hat {p_5}+M_m)];$$
$$A(-,+,-,-,+,+)=Tr\,[\hat {\mu _1}(-\hat {p_4}+m_0)(-1)(1+\gamma _5)\hat {\eta }\hat {k}(-\hat {p_2}+m_0)]$$
$$Tr\,[\hat {\mu _2}(-\hat {q}+m_0)\hat {\mu _1}(\hat {p_1}+m_0)(1+\gamma _5)\hat {k}(\hat {p_3}+m_0)]$$
$$Tr\,[\hat {\mu _2}(-\hat {p_6}+M_m)(-1)(1+\gamma _5)\hat {\eta }\hat {k}(\hat {p_5}+M_m)];$$
$$A(-,+,-,+,-,-)=Tr\,[\hat {\mu _1}(-\hat {p_4}+m_0)(1+\gamma _5)\hat {k}(\hat {p_3}+m_0)$$
$$\hat {\mu _2}(-\hat {q}+m_0)\hat {\mu _1}(\hat {p_1}+m_0)(1-\gamma _5)\hat {k}(-\hat {p_2}+m_0)]$$
$$Tr\,[\hat {\mu _2}(-\hat {p_6}+M_m)(1-\gamma _5)\hat {\eta ^*}\hat {k}(\hat {p_5}+M_m)];$$
$$A(-,+,-,+,-,+)=Tr\,[\hat {\mu _1}(-\hat {p_4}+m_0)(1+\gamma _5)\hat {k}(\hat {p_3}+m_0)$$
$$\hat {\mu _2}(-\hat {q}+m_0)\hat {\mu _1}(\hat {p_1}+m_0)(1-\gamma _5)\hat {k}(-\hat {p_2}+m_0)]$$
$$Tr\,[\hat {\mu _2}(-\hat {p_6}+M_m)(1+\gamma _5)\hat {k}(\hat {p_5}+M_m)];$$
$$A(-,+,-,+,+,-)=Tr\,[\hat {\mu _1}(-\hat {p_4}+m_0)(1+\gamma _5)\hat {k}(\hat {p_3}+m_0)$$
$$\hat {\mu _2}(-\hat {q}+m_0)\hat {\mu _1}(\hat {p_1}+m_0)(1-\gamma _5)\hat {k}(-\hat {p_2}+m_0)]$$
$$Tr\,[\hat {\mu _2}(-\hat {p_6}+M_m)(1-\gamma _5)\hat {k}(\hat {p_5}+M_m)];$$
$$A(-,+,-,+,+,+)=Tr\,[\hat {\mu _1}(-\hat {p_4}+m_0)(1+\gamma _5)\hat {k}(\hat {p_3}+m_0)$$
$$\hat {\mu _2}(-\hat {q}+m_0)\hat {\mu _1}(\hat {p_1}+m_0)(1-\gamma _5)\hat {k}(-\hat {p_2}+m_0)]$$
$$Tr\,[\hat {\mu _2}(-\hat {p_6}+M_m)(-1)(1+\gamma _5)\hat {\eta }\hat {k}(\hat {p_5}+M_m)];$$
$$A(-,+,+,-,-,-)=Tr\,[\hat {\mu _1}(-\hat {p_4}+m_0)(1+\gamma _5)\hat {k}(\hat {p_5}+M_m)$$
$$\hat {\mu _2}(-\hat {p_6}+M_m)(1-\gamma _5)\hat {k}(-\hat {p_2}+m_0)]$$
$$Tr\,[\hat {\mu _2}(-\hat {q}+m_0)\hat {\mu _1}(\hat {p_1}+m_0)(-1)(1+\gamma _5)\hat {\eta }\hat {k}(\hat {p_3}+m_0)];$$
$$A(-,+,+,-,-,+)=Tr\,[\hat {\mu _1}(-\hat {p_4}+m_0)(-1)(1+\gamma _5)\hat {\eta }\hat {k}(-\hat {p_2}+m_0)]$$
$$Tr\,[\hat {\mu _2}(-\hat {q}+m_0)\hat {\mu _1}(\hat {p_1}+m_0)(-1)(1+\gamma _5)\hat {\eta }\hat {k}(\hat {p_3}+m_0)]$$
$$Tr\,[\hat {\mu _2}(-\hat {p_6}+M_m)(1+\gamma _5)\hat {k}(\hat {p_5}+M_m)];$$
$$A(-,+,+,-,+,-)=Tr\,[\hat {\mu _1}(-\hat {p_4}+m_0)(-1)(1+\gamma _5)\hat {\eta }\hat {k}(-\hat {p_2}+m_0)]$$
$$Tr\,[\hat {\mu _2}(-\hat {q}+m_0)\hat {\mu _1}(\hat {p_1}+m_0)(-1)(1+\gamma _5)\hat {\eta }\hat {k}(\hat {p_3}+m_0)]$$
$$Tr\,[\hat {\mu _2}(-\hat {p_6}+M_m)(1-\gamma _5)\hat {k}(\hat {p_5}+M_m)];$$
$$A(-,+,+,-,+,+)=Tr\,[\hat {\mu _1}(-\hat {p_4}+m_0)(-1)(1+\gamma _5)\hat {\eta }\hat {k}(-\hat {p_2}+m_0)]$$
$$Tr\,[\hat {\mu _2}(-\hat {q}+m_0)\hat {\mu _1}(\hat {p_1}+m_0)(-1)(1+\gamma _5)\hat {\eta }\hat {k}(\hat {p_3}+m_0)]$$
$$Tr\,[\hat {\mu _2}(-\hat {p_6}+M_m)(-1)(1+\gamma _5)\hat {\eta }\hat {k}(\hat {p_5}+M_m)];$$
$$A(-,+,+,+,-,-)=Tr\,[\hat {\mu _1}(-\hat {p_4}+m_0)(1+\gamma _5)\hat {k}(\hat {p_5}+M_m)$$
$$\hat {\mu _2}(-\hat {p_6}+M_m)(1-\gamma _5)\hat {k}(-\hat {p_2}+m_0)]$$
$$Tr\,[\hat {\mu _2}(-\hat {q}+m_0)\hat {\mu _1}(\hat {p_1}+m_0)(1-\gamma _5)\hat {k}(\hat {p_3}+m_0)];$$
$$A(-,+,+,+,-,+)=Tr\,[\hat {\mu _1}(-\hat {p_4}+m_0)(-1)(1+\gamma _5)\hat {\eta }\hat {k}(-\hat {p_2}+m_0)]$$
$$Tr\,[\hat {\mu _2}(-\hat {q}+m_0)\hat {\mu _1}(\hat {p_1}+m_0)(1-\gamma _5)\hat {k}(\hat {p_3}+m_0)]$$
$$Tr\,[\hat {\mu _2}(-\hat {p_6}+M_m)(1+\gamma _5)\hat {k}(\hat {p_5}+M_m)];$$
$$A(-,+,+,+,+,-)=Tr\,[\hat {\mu _1}(-\hat {p_4}+m_0)(-1)(1+\gamma _5)\hat {\eta }\hat {k}(-\hat {p_2}+m_0)]$$
$$Tr\,[\hat {\mu _2}(-\hat {q}+m_0)\hat {\mu _1}(\hat {p_1}+m_0)(1-\gamma _5)\hat {k}(\hat {p_3}+m_0)]$$
$$Tr\,[\hat {\mu _2}(-\hat {p_6}+M_m)(1-\gamma _5)\hat {k}(\hat {p_5}+M_m)];$$
$$A(-,+,+,+,+,+)=Tr\,[\hat {\mu _1}(-\hat {p_4}+m_0)(-1)(1+\gamma _5)\hat {\eta }\hat {k}(-\hat {p_2}+m_0)]$$
$$Tr\,[\hat {\mu _2}(-\hat {q}+m_0)\hat {\mu _1}(\hat {p_1}+m_0)(1-\gamma _5)\hat {k}(\hat {p_3}+m_0)]$$
$$Tr\,[\hat {\mu _2}(-\hat {p_6}+M_m)(-1)(1+\gamma _5)\hat {\eta }\hat {k}(\hat {p_5}+M_m)];$$
$$A(+,-,-,-,-,-)=Tr\,[\hat {\mu _1}(-\hat {p_4}+m_0)(1-\gamma _5)\hat {\eta ^*}\hat {k}(-\hat {p_2}+m_0)]$$
$$Tr\,[\hat {\mu _2}(-\hat {q}+m_0)\hat {\mu _1}(\hat {p_1}+m_0)(1+\gamma _5)\hat {k}(\hat {p_3}+m_0)]$$
$$Tr\,[\hat {\mu _2}(-\hat {p_6}+M_m)(1-\gamma _5)\hat {\eta ^*}\hat {k}(\hat {p_5}+M_m)];$$
$$A(+,-,-,-,-,+)=Tr\,[\hat {\mu _1}(-\hat {p_4}+m_0)(1-\gamma _5)\hat {\eta ^*}\hat {k}(-\hat {p_2}+m_0)]$$
$$Tr\,[\hat {\mu _2}(-\hat {q}+m_0)\hat {\mu _1}(\hat {p_1}+m_0)(1+\gamma _5)\hat {k}(\hat {p_3}+m_0)]$$
$$Tr\,[\hat {\mu _2}(-\hat {p_6}+M_m)(1+\gamma _5)\hat {k}(\hat {p_5}+M_m)];$$
$$A(+,-,-,-,+,-)=Tr\,[\hat {\mu _1}(-\hat {p_4}+m_0)(1-\gamma _5)\hat {\eta ^*}\hat {k}(-\hat {p_2}+m_0)]$$
$$Tr\,[\hat {\mu _2}(-\hat {q}+m_0)\hat {\mu _1}(\hat {p_1}+m_0)(1+\gamma _5)\hat {k}(\hat {p_3}+m_0)]$$
$$Tr\,[\hat {\mu _2}(-\hat {p_6}+M_m)(1-\gamma _5)\hat {k}(\hat {p_5}+M_m)];$$
$$A(+,-,-,-,+,+)=Tr\,[\hat {\mu _1}(-\hat {p_4}+m_0)(1-\gamma _5)\hat {k}(\hat {p_5}+M_m)$$
$$\hat {\mu _2}(-\hat {p_6}+M_m)(1+\gamma _5)\hat {k}(-\hat {p_2}+m_0)]$$
$$Tr\,[\hat {\mu _2}(-\hat {q}+m_0)\hat {\mu _1}(\hat {p_1}+m_0)(1+\gamma _5)\hat {k}(\hat {p_3}+m_0)];$$
$$A(+,-,-,+,-,-)=Tr\,[\hat {\mu _1}(-\hat {p_4}+m_0)(1-\gamma _5)\hat {\eta ^*}\hat {k}(-\hat {p_2}+m_0)]$$
$$Tr\,[\hat {\mu _2}(-\hat {q}+m_0)\hat {\mu _1}(\hat {p_1}+m_0)(1-\gamma _5)\hat {\eta ^*}\hat {k}(\hat {p_3}+m_0)]$$
$$Tr\,[\hat {\mu _2}(-\hat {p_6}+M_m)(1-\gamma _5)\hat {\eta ^*}\hat {k}(\hat {p_5}+M_m)];$$
$$A(+,-,-,+,-,+)=Tr\,[\hat {\mu _1}(-\hat {p_4}+m_0)(1-\gamma _5)\hat {\eta ^*}\hat {k}(-\hat {p_2}+m_0)]$$
$$Tr\,[\hat {\mu _2}(-\hat {q}+m_0)\hat {\mu _1}(\hat {p_1}+m_0)(1-\gamma _5)\hat {\eta ^*}\hat {k}(\hat {p_3}+m_0)]$$
$$Tr\,[\hat {\mu _2}(-\hat {p_6}+M_m)(1+\gamma _5)\hat {k}(\hat {p_5}+M_m)];$$
$$A(+,-,-,+,+,-)=Tr\,[\hat {\mu _1}(-\hat {p_4}+m_0)(1-\gamma _5)\hat {\eta ^*}\hat {k}(-\hat {p_2}+m_0)]$$
$$Tr\,[\hat {\mu _2}(-\hat {q}+m_0)\hat {\mu _1}(\hat {p_1}+m_0)(1-\gamma _5)\hat {\eta ^*}\hat {k}(\hat {p_3}+m_0)]$$
$$Tr\,[\hat {\mu _2}(-\hat {p_6}+M_m)(1-\gamma _5)\hat {k}(\hat {p_5}+M_m)];$$
$$A(+,-,-,+,+,+)=Tr\,[\hat {\mu _1}(-\hat {p_4}+m_0)(1-\gamma _5)\hat {k}(\hat {p_5}+M_m)$$
$$\hat {\mu _2}(-\hat {p_6}+M_m)(1+\gamma _5)\hat {k}(-\hat {p_2}+m_0)]$$
$$Tr\,[\hat {\mu _2}(-\hat {q}+m_0)\hat {\mu _1}(\hat {p_1}+m_0)(1-\gamma _5)\hat {\eta ^*}\hat {k}(\hat {p_3}+m_0)];$$
$$A(+,-,+,-,-,-)=Tr\,[\hat {\mu _1}(-\hat {p_4}+m_0)(1-\gamma _5)\hat {k}(\hat {p_3}+m_0)$$
$$\hat {\mu _2}(-\hat {q}+m_0)\hat {\mu _1}(\hat {p_1}+m_0)(1+\gamma _5)\hat {k}(-\hat {p_2}+m_0)]$$
$$Tr\,[\hat {\mu _2}(-\hat {p_6}+M_m)(1-\gamma _5)\hat {\eta ^*}\hat {k}(\hat {p_5}+M_m)];$$
$$A(+,-,+,-,-,+)=Tr\,[\hat {\mu _1}(-\hat {p_4}+m_0)(1-\gamma _5)\hat {k}(\hat {p_3}+m_0)$$
$$\hat {\mu _2}(-\hat {q}+m_0)\hat {\mu _1}(\hat {p_1}+m_0)(1+\gamma _5)\hat {k}(-\hat {p_2}+m_0)]$$
$$Tr\,[\hat {\mu _2}(-\hat {p_6}+M_m)(1+\gamma _5)\hat {k}(\hat {p_5}+M_m)];$$
$$A(+,-,+,-,+,-)=Tr\,[\hat {\mu _1}(-\hat {p_4}+m_0)(1-\gamma _5)\hat {k}(\hat {p_3}+m_0)$$
$$\hat {\mu _2}(-\hat {q}+m_0)\hat {\mu _1}(\hat {p_1}+m_0)(1+\gamma _5)\hat {k}(-\hat {p_2}+m_0)]$$
$$Tr\,[\hat {\mu _2}(-\hat {p_6}+M_m)(1-\gamma _5)\hat {k}(\hat {p_5}+M_m)];$$
$$A(+,-,+,-,+,+)=Tr\,[\hat {\mu _1}(-\hat {p_4}+m_0)(1-\gamma _5)\hat {k}(\hat {p_3}+m_0)$$
$$\hat {\mu _2}(-\hat {q}+m_0)\hat {\mu _1}(\hat {p_1}+m_0)(1+\gamma _5)\hat {k}(-\hat {p_2}+m_0)]$$
$$Tr\,[\hat {\mu _2}(-\hat {p_6}+M_m)(-1)(1+\gamma _5)\hat {\eta }\hat {k}(\hat {p_5}+M_m)];$$
$$A(+,-,+,+,-,-)=Tr\,[\hat {\mu _1}(-\hat {p_4}+m_0)(1-\gamma _5)\hat {\eta ^*}\hat {k}(-\hat {p_2}+m_0)]$$
$$Tr\,[\hat {\mu _2}(-\hat {q}+m_0)\hat {\mu _1}(\hat {p_1}+m_0)(1-\gamma _5)\hat {k}(\hat {p_3}+m_0)]$$
$$Tr\,[\hat {\mu _2}(-\hat {p_6}+M_m)(1-\gamma _5)\hat {\eta ^*}\hat {k}(\hat {p_5}+M_m)];$$
$$A(+,-,+,+,-,+)=Tr\,[\hat {\mu _1}(-\hat {p_4}+m_0)(1-\gamma _5)\hat {\eta ^*}\hat {k}(-\hat {p_2}+m_0)]$$
$$Tr\,[\hat {\mu _2}(-\hat {q}+m_0)\hat {\mu _1}(\hat {p_1}+m_0)(1-\gamma _5)\hat {k}(\hat {p_3}+m_0)]$$
$$Tr\,[\hat {\mu _2}(-\hat {p_6}+M_m)(1+\gamma _5)\hat {k}(\hat {p_5}+M_m)];$$
$$A(+,-,+,+,+,-)=Tr\,[\hat {\mu _1}(-\hat {p_4}+m_0)(1-\gamma _5)\hat {\eta ^*}\hat {k}(-\hat {p_2}+m_0)]$$
$$Tr\,[\hat {\mu _2}(-\hat {q}+m_0)\hat {\mu _1}(\hat {p_1}+m_0)(1-\gamma _5)\hat {k}(\hat {p_3}+m_0)]$$
$$Tr\,[\hat {\mu _2}(-\hat {p_6}+M_m)(1-\gamma _5)\hat {k}(\hat {p_5}+M_m)];$$
$$A(+,-,+,+,+,+)=Tr\,[\hat {\mu _1}(-\hat {p_4}+m_0)(1-\gamma _5)\hat {k}(\hat {p_5}+M_m)$$
$$\hat {\mu _2}(-\hat {p_6}+M_m)(1+\gamma _5)\hat {k}(-\hat {p_2}+m_0)]$$
$$Tr\,[\hat {\mu _2}(-\hat {q}+m_0)\hat {\mu _1}(\hat {p_1}+m_0)(1-\gamma _5)\hat {k}(\hat {p_3}+m_0)];$$
$$A(+,+,-,-,-,-)=Tr\,[\hat {\mu _1}(-\hat {p_4}+m_0)(1+\gamma _5)\hat {k}(-\hat {p_2}+m_0)]$$
$$Tr\,[\hat {\mu _2}(-\hat {q}+m_0)\hat {\mu _1}(\hat {p_1}+m_0)(1+\gamma _5)\hat {k}(\hat {p_3}+m_0)]$$
$$Tr\,[\hat {\mu _2}(-\hat {p_6}+M_m)(1-\gamma _5)\hat {\eta ^*}\hat {k}(\hat {p_5}+M_m)];$$
$$A(+,+,-,-,-,+)=Tr\,[\hat {\mu _1}(-\hat {p_4}+m_0)(1+\gamma _5)\hat {k}(-\hat {p_2}+m_0)]$$
$$Tr\,[\hat {\mu _2}(-\hat {q}+m_0)\hat {\mu _1}(\hat {p_1}+m_0)(1+\gamma _5)\hat {k}(\hat {p_3}+m_0)]$$
$$Tr\,[\hat {\mu _2}(-\hat {p_6}+M_m)(1+\gamma _5)\hat {k}(\hat {p_5}+M_m)];$$
$$A(+,+,-,-,+,-)=Tr\,[\hat {\mu _1}(-\hat {p_4}+m_0)(1+\gamma _5)\hat {k}(-\hat {p_2}+m_0)]$$
$$Tr\,[\hat {\mu _2}(-\hat {q}+m_0)\hat {\mu _1}(\hat {p_1}+m_0)(1+\gamma _5)\hat {k}(\hat {p_3}+m_0)]$$
$$Tr\,[\hat {\mu _2}(-\hat {p_6}+M_m)(1-\gamma _5)\hat {k}(\hat {p_5}+M_m)];$$
$$A(+,+,-,-,+,+)=Tr\,[\hat {\mu _1}(-\hat {p_4}+m_0)(1+\gamma _5)\hat {k}(-\hat {p_2}+m_0)]$$
$$Tr\,[\hat {\mu _2}(-\hat {q}+m_0)\hat {\mu _1}(\hat {p_1}+m_0)(1+\gamma _5)\hat {k}(\hat {p_3}+m_0)]$$
$$Tr\,[\hat {\mu _2}(-\hat {p_6}+M_m)(-1)(1+\gamma _5)\hat {\eta }\hat {k}(\hat {p_5}+M_m)];$$
$$A(+,+,-,+,-,-)=Tr\,[\hat {\mu _1}(-\hat {p_4}+m_0)(1+\gamma _5)\hat {k}(-\hat {p_2}+m_0)]$$
$$Tr\,[\hat {\mu _2}(-\hat {q}+m_0)\hat {\mu _1}(\hat {p_1}+m_0)(1-\gamma _5)\hat {\eta ^*}\hat {k}(\hat {p_3}+m_0)]$$
$$Tr\,[\hat {\mu _2}(-\hat {p_6}+M_m)(1-\gamma _5)\hat {\eta ^*}\hat {k}(\hat {p_5}+M_m)];$$
$$A(+,+,-,+,-,+)=Tr\,[\hat {\mu _1}(-\hat {p_4}+m_0)(1+\gamma _5)\hat {k}(-\hat {p_2}+m_0)]$$
$$Tr\,[\hat {\mu _2}(-\hat {q}+m_0)\hat {\mu _1}(\hat {p_1}+m_0)(1-\gamma _5)\hat {\eta ^*}\hat {k}(\hat {p_3}+m_0)]$$
$$Tr\,[\hat {\mu _2}(-\hat {p_6}+M_m)(1+\gamma _5)\hat {k}(\hat {p_5}+M_m)];$$
$$A(+,+,-,+,+,-)=Tr\,[\hat {\mu _1}(-\hat {p_4}+m_0)(1+\gamma _5)\hat {k}(-\hat {p_2}+m_0)]$$
$$Tr\,[\hat {\mu _2}(-\hat {q}+m_0)\hat {\mu _1}(\hat {p_1}+m_0)(1-\gamma _5)\hat {\eta ^*}\hat {k}(\hat {p_3}+m_0)]$$
$$Tr\,[\hat {\mu _2}(-\hat {p_6}+M_m)(1-\gamma _5)\hat {k}(\hat {p_5}+M_m)];$$
$$A(+,+,-,+,+,+)=Tr\,[\hat {\mu _1}(-\hat {p_4}+m_0)(1+\gamma _5)\hat {k}(-\hat {p_2}+m_0)]$$
$$Tr\,[\hat {\mu _2}(-\hat {q}+m_0)\hat {\mu _1}(\hat {p_1}+m_0)(1-\gamma _5)\hat {k}(\hat {p_5}+M_m)$$
$$\hat {\mu _2}(-\hat {p_6}+M_m)(1+\gamma _5)\hat {k}(\hat {p_3}+m_0)];$$
$$A(+,+,+,-,-,-)=Tr\,[\hat {\mu _1}(-\hat {p_4}+m_0)(1+\gamma _5)\hat {k}(-\hat {p_2}+m_0)]$$
$$Tr\,[\hat {\mu _2}(-\hat {q}+m_0)\hat {\mu _1}(\hat {p_1}+m_0)(1+\gamma _5)\hat {k}(\hat {p_5}+M_m)$$
$$\hat {\mu _2}(-\hat {p_6}+M_m)(1-\gamma _5)\hat {k}(\hat {p_3}+m_0)];$$
$$A(+,+,+,-,-,+)=Tr\,[\hat {\mu _1}(-\hat {p_4}+m_0)(1+\gamma _5)\hat {k}(-\hat {p_2}+m_0)]$$
$$Tr\,[\hat {\mu _2}(-\hat {q}+m_0)\hat {\mu _1}(\hat {p_1}+m_0)(-1)(1+\gamma _5)\hat {\eta }\hat {k}(\hat {p_3}+m_0)]$$
$$Tr\,[\hat {\mu _2}(-\hat {p_6}+M_m)(1+\gamma _5)\hat {k}(\hat {p_5}+M_m)];$$
$$A(+,+,+,-,+,-)=Tr\,[\hat {\mu _1}(-\hat {p_4}+m_0)(1+\gamma _5)\hat {k}(-\hat {p_2}+m_0)]$$
$$Tr\,[\hat {\mu _2}(-\hat {q}+m_0)\hat {\mu _1}(\hat {p_1}+m_0)(-1)(1+\gamma _5)\hat {\eta }\hat {k}(\hat {p_3}+m_0)]$$
$$Tr\,[\hat {\mu _2}(-\hat {p_6}+M_m)(1-\gamma _5)\hat {k}(\hat {p_5}+M_m)];$$
$$A(+,+,+,-,+,+)=Tr\,[\hat {\mu _1}(-\hat {p_4}+m_0)(1+\gamma _5)\hat {k}(-\hat {p_2}+m_0)]$$
$$Tr\,[\hat {\mu _2}(-\hat {q}+m_0)\hat {\mu _1}(\hat {p_1}+m_0)(-1)(1+\gamma _5)\hat {\eta }\hat {k}(\hat {p_3}+m_0)]$$
$$Tr\,[\hat {\mu _2}(-\hat {p_6}+M_m)(-1)(1+\gamma _5)\hat {\eta }\hat {k}(\hat {p_5}+M_m)];$$
$$A(+,+,+,+,-,-)=Tr\,[\hat {\mu _1}(-\hat {p_4}+m_0)(1+\gamma _5)\hat {k}(-\hat {p_2}+m_0)]$$
$$Tr\,[\hat {\mu _2}(-\hat {q}+m_0)\hat {\mu _1}(\hat {p_1}+m_0)(1-\gamma _5)\hat {k}(\hat {p_3}+m_0)]$$
$$Tr\,[\hat {\mu _2}(-\hat {p_6}+M_m)(1-\gamma _5)\hat {\eta ^*}\hat {k}(\hat {p_5}+M_m)];$$
$$A(+,+,+,+,-,+)=Tr\,[\hat {\mu _1}(-\hat {p_4}+m_0)(1+\gamma _5)\hat {k}(-\hat {p_2}+m_0)]$$
$$Tr\,[\hat {\mu _2}(-\hat {q}+m_0)\hat {\mu _1}(\hat {p_1}+m_0)(1-\gamma _5)\hat {k}(\hat {p_3}+m_0)]$$
$$Tr\,[\hat {\mu _2}(-\hat {p_6}+M_m)(1+\gamma _5)\hat {k}(\hat {p_5}+M_m)];$$
$$A(+,+,+,+,+,-)=Tr\,[\hat {\mu _1}(-\hat {p_4}+m_0)(1+\gamma _5)\hat {k}(-\hat {p_2}+m_0)]$$
$$Tr\,[\hat {\mu _2}(-\hat {q}+m_0)\hat {\mu _1}(\hat {p_1}+m_0)(1-\gamma _5)\hat {k}(\hat {p_3}+m_0)]$$
$$Tr\,[\hat {\mu _2}(-\hat {p_6}+M_m)(1-\gamma _5)\hat {k}(\hat {p_5}+M_m)];$$
$$A(+,+,+,+,+,+)=Tr\,[\hat {\mu _1}(-\hat {p_4}+m_0)(1+\gamma _5)\hat {k}(-\hat {p_2}+m_0)]$$
$$Tr\,[\hat {\mu _2}(-\hat {q}+m_0)\hat {\mu _1}(\hat {p_1}+m_0)(1-\gamma _5)\hat {k}(\hat {p_3}+m_0)]$$
$$Tr\,[\hat {\mu _2}(-\hat {p_6}+M_m)(-1)(1+\gamma _5)\hat {\eta }\hat {k}(\hat {p_5}+M_m)].$$

\unitlength=1.0 pt
\SetScale{1.0}
\SetWidth{0.7}      
\tiny    
{} \qquad\allowbreak

\noindent
\begin{picture}(79,81)(0,0)
\Text(13.0,65.0)[r]{$e$}
\ArrowLine(14.0,65.0)(31.0,65.0) 
\Text(39.0,69.0)[b]{$e$}
\ArrowLine(31.0,65.0)(48.0,65.0) 
\Text(66.0,73.0)[l]{$e$}
\ArrowLine(48.0,65.0)(65.0,73.0) 
\Text(47.0,57.0)[r]{$\gamma$}
\DashLine(48.0,65.0)(48.0,49.0){3.0} 
\Text(66.0,57.0)[l]{$\mu$}
\ArrowLine(48.0,49.0)(65.0,57.0) 
\Text(66.0,41.0)[l]{$\bar{\mu}$}
\ArrowLine(65.0,41.0)(48.0,49.0) 
\Text(30.0,49.0)[r]{$\gamma$}
\DashLine(31.0,65.0)(31.0,33.0){3.0} 
\Text(13.0,33.0)[r]{$\bar{e}$}
\ArrowLine(31.0,33.0)(14.0,33.0) 
\Line(31.0,33.0)(48.0,33.0) 
\Text(66.0,25.0)[l]{$\bar{e}$}
\ArrowLine(65.0,25.0)(48.0,33.0) 
\Text(39,0)[b] {diagr.1}
\end{picture} \ 
{} \qquad\allowbreak
\begin{picture}(79,81)(0,0)
\Text(13.0,65.0)[r]{$e$}
\ArrowLine(14.0,65.0)(31.0,65.0) 
\Text(39.0,66.0)[b]{$\gamma$}
\DashLine(31.0,65.0)(48.0,65.0){3.0} 
\Text(66.0,73.0)[l]{$e$}
\ArrowLine(48.0,65.0)(65.0,73.0) 
\Text(66.0,57.0)[l]{$\bar{e}$}
\ArrowLine(65.0,57.0)(48.0,65.0) 
\Text(27.0,49.0)[r]{$e$}
\ArrowLine(31.0,65.0)(31.0,33.0) 
\Text(13.0,33.0)[r]{$\bar{e}$}
\ArrowLine(31.0,33.0)(14.0,33.0) 
\Text(39.0,34.0)[b]{$\gamma$}
\DashLine(31.0,33.0)(48.0,33.0){3.0} 
\Text(66.0,41.0)[l]{$\mu$}
\ArrowLine(48.0,33.0)(65.0,41.0) 
\Text(66.0,25.0)[l]{$\bar{\mu}$}
\ArrowLine(65.0,25.0)(48.0,33.0) 
\Text(39,0)[b] {diagr.2}
\end{picture} \ 
{} \qquad\allowbreak
\begin{picture}(79,81)(0,0)
\Text(13.0,65.0)[r]{$e$}
\ArrowLine(14.0,65.0)(31.0,65.0) 
\Text(39.0,69.0)[b]{$e$}
\ArrowLine(31.0,65.0)(48.0,65.0) 
\Text(66.0,73.0)[l]{$e$}
\ArrowLine(48.0,65.0)(65.0,73.0) 
\Text(47.0,57.0)[r]{$Z$}
\DashLine(48.0,65.0)(48.0,49.0){3.0} 
\Text(66.0,57.0)[l]{$\mu$}
\ArrowLine(48.0,49.0)(65.0,57.0) 
\Text(66.0,41.0)[l]{$\bar{\mu}$}
\ArrowLine(65.0,41.0)(48.0,49.0) 
\Text(30.0,49.0)[r]{$\gamma$}
\DashLine(31.0,65.0)(31.0,33.0){3.0} 
\Text(13.0,33.0)[r]{$\bar{e}$}
\ArrowLine(31.0,33.0)(14.0,33.0) 
\Line(31.0,33.0)(48.0,33.0) 
\Text(66.0,25.0)[l]{$\bar{e}$}
\ArrowLine(65.0,25.0)(48.0,33.0) 
\Text(39,0)[b] {diagr.3}
\end{picture} \ 
{} \qquad\allowbreak
\begin{picture}(79,81)(0,0)
\Text(13.0,65.0)[r]{$e$}
\ArrowLine(14.0,65.0)(31.0,65.0) 
\Text(39.0,66.0)[b]{$\gamma$}
\DashLine(31.0,65.0)(48.0,65.0){3.0} 
\Text(66.0,73.0)[l]{$e$}
\ArrowLine(48.0,65.0)(65.0,73.0) 
\Text(66.0,57.0)[l]{$\bar{e}$}
\ArrowLine(65.0,57.0)(48.0,65.0) 
\Text(27.0,49.0)[r]{$e$}
\ArrowLine(31.0,65.0)(31.0,33.0) 
\Text(13.0,33.0)[r]{$\bar{e}$}
\ArrowLine(31.0,33.0)(14.0,33.0) 
\Text(39.0,34.0)[b]{$Z$}
\DashLine(31.0,33.0)(48.0,33.0){3.0} 
\Text(66.0,41.0)[l]{$\mu$}
\ArrowLine(48.0,33.0)(65.0,41.0) 
\Text(66.0,25.0)[l]{$\bar{\mu}$}
\ArrowLine(65.0,25.0)(48.0,33.0) 
\Text(39,0)[b] {diagr.4}
\end{picture} \ 
{} \qquad\allowbreak
\begin{picture}(79,81)(0,0)
\Text(13.0,65.0)[r]{$e$}
\ArrowLine(14.0,65.0)(31.0,65.0) 
\Text(39.0,66.0)[b]{$\gamma$}
\DashLine(31.0,65.0)(48.0,65.0){3.0} 
\Text(66.0,73.0)[l]{$\mu$}
\ArrowLine(48.0,65.0)(65.0,73.0) 
\Text(66.0,57.0)[l]{$\bar{\mu}$}
\ArrowLine(65.0,57.0)(48.0,65.0) 
\Text(27.0,49.0)[r]{$e$}
\ArrowLine(31.0,65.0)(31.0,33.0) 
\Text(13.0,33.0)[r]{$\bar{e}$}
\ArrowLine(31.0,33.0)(14.0,33.0) 
\Text(39.0,34.0)[b]{$\gamma$}
\DashLine(31.0,33.0)(48.0,33.0){3.0} 
\Text(66.0,41.0)[l]{$e$}
\ArrowLine(48.0,33.0)(65.0,41.0) 
\Text(66.0,25.0)[l]{$\bar{e}$}
\ArrowLine(65.0,25.0)(48.0,33.0) 
\Text(39,0)[b] {diagr.5}
\end{picture} \ 
{} \qquad\allowbreak
\begin{picture}(79,81)(0,0)
\Text(13.0,65.0)[r]{$e$}
\ArrowLine(14.0,65.0)(31.0,65.0) 
\Text(39.0,66.0)[b]{$\gamma$}
\DashLine(31.0,65.0)(48.0,65.0){3.0} 
\Text(66.0,73.0)[l]{$\mu$}
\ArrowLine(48.0,65.0)(65.0,73.0) 
\Text(66.0,57.0)[l]{$\bar{\mu}$}
\ArrowLine(65.0,57.0)(48.0,65.0) 
\Text(27.0,57.0)[r]{$e$}
\ArrowLine(31.0,65.0)(31.0,49.0) 
\Line(31.0,49.0)(48.0,49.0) 
\Text(66.0,41.0)[l]{$e$}
\ArrowLine(48.0,49.0)(65.0,41.0) 
\Text(30.0,41.0)[r]{$\gamma$}
\DashLine(31.0,49.0)(31.0,33.0){3.0} 
\Text(13.0,33.0)[r]{$\bar{e}$}
\ArrowLine(31.0,33.0)(14.0,33.0) 
\Line(31.0,33.0)(48.0,33.0) 
\Text(66.0,25.0)[l]{$\bar{e}$}
\ArrowLine(65.0,25.0)(48.0,33.0) 
\Text(39,0)[b] {diagr.6}
\end{picture} \ 
{} \qquad\allowbreak
\begin{picture}(79,81)(0,0)
\Text(13.0,65.0)[r]{$e$}
\ArrowLine(14.0,65.0)(31.0,65.0) 
\Text(39.0,66.0)[b]{$\gamma$}
\DashLine(31.0,65.0)(48.0,65.0){3.0} 
\Text(66.0,73.0)[l]{$\mu$}
\ArrowLine(48.0,65.0)(65.0,73.0) 
\Text(66.0,57.0)[l]{$\bar{\mu}$}
\ArrowLine(65.0,57.0)(48.0,65.0) 
\Text(27.0,49.0)[r]{$e$}
\ArrowLine(31.0,65.0)(31.0,33.0) 
\Text(13.0,33.0)[r]{$\bar{e}$}
\ArrowLine(31.0,33.0)(14.0,33.0) 
\Text(39.0,34.0)[b]{$Z$}
\DashLine(31.0,33.0)(48.0,33.0){3.0} 
\Text(66.0,41.0)[l]{$e$}
\ArrowLine(48.0,33.0)(65.0,41.0) 
\Text(66.0,25.0)[l]{$\bar{e}$}
\ArrowLine(65.0,25.0)(48.0,33.0) 
\Text(39,0)[b] {diagr.7}
\end{picture} \ 
{} \qquad\allowbreak
\begin{picture}(79,81)(0,0)
\Text(13.0,65.0)[r]{$e$}
\ArrowLine(14.0,65.0)(31.0,65.0) 
\Text(39.0,66.0)[b]{$\gamma$}
\DashLine(31.0,65.0)(48.0,65.0){3.0} 
\Text(66.0,73.0)[l]{$\mu$}
\ArrowLine(48.0,65.0)(65.0,73.0) 
\Text(66.0,57.0)[l]{$\bar{\mu}$}
\ArrowLine(65.0,57.0)(48.0,65.0) 
\Text(27.0,57.0)[r]{$e$}
\ArrowLine(31.0,65.0)(31.0,49.0) 
\Line(31.0,49.0)(48.0,49.0) 
\Text(66.0,41.0)[l]{$e$}
\ArrowLine(48.0,49.0)(65.0,41.0) 
\Text(30.0,41.0)[r]{$Z$}
\DashLine(31.0,49.0)(31.0,33.0){3.0} 
\Text(13.0,33.0)[r]{$\bar{e}$}
\ArrowLine(31.0,33.0)(14.0,33.0) 
\Line(31.0,33.0)(48.0,33.0) 
\Text(66.0,25.0)[l]{$\bar{e}$}
\ArrowLine(65.0,25.0)(48.0,33.0) 
\Text(39,0)[b] {diagr.8}
\end{picture} \ 
{} \qquad\allowbreak
\begin{picture}(79,81)(0,0)
\Text(13.0,73.0)[r]{$e$}
\ArrowLine(14.0,73.0)(31.0,65.0) 
\Text(13.0,57.0)[r]{$\bar{e}$}
\ArrowLine(31.0,65.0)(14.0,57.0) 
\Text(39.0,66.0)[b]{$\gamma$}
\DashLine(31.0,65.0)(48.0,65.0){3.0} 
\Text(66.0,73.0)[l]{$\bar{e}$}
\ArrowLine(65.0,73.0)(48.0,65.0) 
\Text(44.0,57.0)[r]{$e$}
\ArrowLine(48.0,65.0)(48.0,49.0) 
\Text(66.0,57.0)[l]{$e$}
\ArrowLine(48.0,49.0)(65.0,57.0) 
\Text(47.0,41.0)[r]{$\gamma$}
\DashLine(48.0,49.0)(48.0,33.0){3.0} 
\Text(66.0,41.0)[l]{$\mu$}
\ArrowLine(48.0,33.0)(65.0,41.0) 
\Text(66.0,25.0)[l]{$\bar{\mu}$}
\ArrowLine(65.0,25.0)(48.0,33.0) 
\Text(39,0)[b] {diagr.9}
\end{picture} \ 
{} \qquad\allowbreak
\begin{picture}(79,81)(0,0)
\Text(13.0,65.0)[r]{$e$}
\ArrowLine(14.0,65.0)(31.0,65.0) 
\Line(31.0,65.0)(48.0,65.0) 
\Text(66.0,73.0)[l]{$e$}
\ArrowLine(48.0,65.0)(65.0,73.0) 
\Text(30.0,57.0)[r]{$\gamma$}
\DashLine(31.0,65.0)(31.0,49.0){3.0} 
\Line(31.0,49.0)(48.0,49.0) 
\Text(66.0,57.0)[l]{$\bar{e}$}
\ArrowLine(65.0,57.0)(48.0,49.0) 
\Text(27.0,41.0)[r]{$e$}
\ArrowLine(31.0,49.0)(31.0,33.0) 
\Text(13.0,33.0)[r]{$\bar{e}$}
\ArrowLine(31.0,33.0)(14.0,33.0) 
\Text(39.0,34.0)[b]{$\gamma$}
\DashLine(31.0,33.0)(48.0,33.0){3.0} 
\Text(66.0,41.0)[l]{$\mu$}
\ArrowLine(48.0,33.0)(65.0,41.0) 
\Text(66.0,25.0)[l]{$\bar{\mu}$}
\ArrowLine(65.0,25.0)(48.0,33.0) 
\Text(39,0)[b] {diagr.10}
\end{picture} \ 
{} \qquad\allowbreak
\begin{picture}(79,81)(0,0)
\Text(13.0,73.0)[r]{$e$}
\ArrowLine(14.0,73.0)(31.0,65.0) 
\Text(13.0,57.0)[r]{$\bar{e}$}
\ArrowLine(31.0,65.0)(14.0,57.0) 
\Text(39.0,66.0)[b]{$\gamma$}
\DashLine(31.0,65.0)(48.0,65.0){3.0} 
\Text(66.0,73.0)[l]{$\bar{e}$}
\ArrowLine(65.0,73.0)(48.0,65.0) 
\Text(44.0,57.0)[r]{$e$}
\ArrowLine(48.0,65.0)(48.0,49.0) 
\Text(66.0,57.0)[l]{$e$}
\ArrowLine(48.0,49.0)(65.0,57.0) 
\Text(47.0,41.0)[r]{$Z$}
\DashLine(48.0,49.0)(48.0,33.0){3.0} 
\Text(66.0,41.0)[l]{$\mu$}
\ArrowLine(48.0,33.0)(65.0,41.0) 
\Text(66.0,25.0)[l]{$\bar{\mu}$}
\ArrowLine(65.0,25.0)(48.0,33.0) 
\Text(39,0)[b] {diagr.11}
\end{picture} \ 
{} \qquad\allowbreak
\begin{picture}(79,81)(0,0)
\Text(13.0,65.0)[r]{$e$}
\ArrowLine(14.0,65.0)(31.0,65.0) 
\Line(31.0,65.0)(48.0,65.0) 
\Text(66.0,73.0)[l]{$e$}
\ArrowLine(48.0,65.0)(65.0,73.0) 
\Text(30.0,57.0)[r]{$\gamma$}
\DashLine(31.0,65.0)(31.0,49.0){3.0} 
\Line(31.0,49.0)(48.0,49.0) 
\Text(66.0,57.0)[l]{$\bar{e}$}
\ArrowLine(65.0,57.0)(48.0,49.0) 
\Text(27.0,41.0)[r]{$e$}
\ArrowLine(31.0,49.0)(31.0,33.0) 
\Text(13.0,33.0)[r]{$\bar{e}$}
\ArrowLine(31.0,33.0)(14.0,33.0) 
\Text(39.0,34.0)[b]{$Z$}
\DashLine(31.0,33.0)(48.0,33.0){3.0} 
\Text(66.0,41.0)[l]{$\mu$}
\ArrowLine(48.0,33.0)(65.0,41.0) 
\Text(66.0,25.0)[l]{$\bar{\mu}$}
\ArrowLine(65.0,25.0)(48.0,33.0) 
\Text(39,0)[b] {diagr.12}
\end{picture} \ 
{} \qquad\allowbreak
\begin{picture}(79,81)(0,0)
\Text(13.0,73.0)[r]{$e$}
\ArrowLine(14.0,73.0)(31.0,65.0) 
\Text(13.0,57.0)[r]{$\bar{e}$}
\ArrowLine(31.0,65.0)(14.0,57.0) 
\Text(39.0,66.0)[b]{$\gamma$}
\DashLine(31.0,65.0)(48.0,65.0){3.0} 
\Text(66.0,73.0)[l]{$e$}
\ArrowLine(48.0,65.0)(65.0,73.0) 
\Text(44.0,57.0)[r]{$e$}
\ArrowLine(48.0,49.0)(48.0,65.0) 
\Text(66.0,57.0)[l]{$\bar{e}$}
\ArrowLine(65.0,57.0)(48.0,49.0) 
\Text(47.0,41.0)[r]{$\gamma$}
\DashLine(48.0,49.0)(48.0,33.0){3.0} 
\Text(66.0,41.0)[l]{$\mu$}
\ArrowLine(48.0,33.0)(65.0,41.0) 
\Text(66.0,25.0)[l]{$\bar{\mu}$}
\ArrowLine(65.0,25.0)(48.0,33.0) 
\Text(39,0)[b] {diagr.13}
\end{picture} \ 
{} \qquad\allowbreak
\begin{picture}(79,81)(0,0)
\Text(13.0,65.0)[r]{$e$}
\ArrowLine(14.0,65.0)(31.0,65.0) 
\Line(31.0,65.0)(48.0,65.0) 
\Text(66.0,73.0)[l]{$e$}
\ArrowLine(48.0,65.0)(65.0,73.0) 
\Text(30.0,57.0)[r]{$\gamma$}
\DashLine(31.0,65.0)(31.0,49.0){3.0} 
\Text(13.0,49.0)[r]{$\bar{e}$}
\ArrowLine(31.0,49.0)(14.0,49.0) 
\Text(39.0,53.0)[b]{$e$}
\ArrowLine(48.0,49.0)(31.0,49.0) 
\Text(66.0,57.0)[l]{$\bar{e}$}
\ArrowLine(65.0,57.0)(48.0,49.0) 
\Text(47.0,41.0)[r]{$\gamma$}
\DashLine(48.0,49.0)(48.0,33.0){3.0} 
\Text(66.0,41.0)[l]{$\mu$}
\ArrowLine(48.0,33.0)(65.0,41.0) 
\Text(66.0,25.0)[l]{$\bar{\mu}$}
\ArrowLine(65.0,25.0)(48.0,33.0) 
\Text(39,0)[b] {diagr.14}
\end{picture} \ 
{} \qquad\allowbreak
\begin{picture}(79,81)(0,0)
\Text(13.0,73.0)[r]{$e$}
\ArrowLine(14.0,73.0)(31.0,65.0) 
\Text(13.0,57.0)[r]{$\bar{e}$}
\ArrowLine(31.0,65.0)(14.0,57.0) 
\Text(39.0,66.0)[b]{$\gamma$}
\DashLine(31.0,65.0)(48.0,65.0){3.0} 
\Text(66.0,73.0)[l]{$e$}
\ArrowLine(48.0,65.0)(65.0,73.0) 
\Text(44.0,57.0)[r]{$e$}
\ArrowLine(48.0,49.0)(48.0,65.0) 
\Text(66.0,57.0)[l]{$\bar{e}$}
\ArrowLine(65.0,57.0)(48.0,49.0) 
\Text(47.0,41.0)[r]{$Z$}
\DashLine(48.0,49.0)(48.0,33.0){3.0} 
\Text(66.0,41.0)[l]{$\mu$}
\ArrowLine(48.0,33.0)(65.0,41.0) 
\Text(66.0,25.0)[l]{$\bar{\mu}$}
\ArrowLine(65.0,25.0)(48.0,33.0) 
\Text(39,0)[b] {diagr.15}
\end{picture} \ 
{} \qquad\allowbreak
\begin{picture}(79,81)(0,0)
\Text(13.0,65.0)[r]{$e$}
\ArrowLine(14.0,65.0)(31.0,65.0) 
\Line(31.0,65.0)(48.0,65.0) 
\Text(66.0,73.0)[l]{$e$}
\ArrowLine(48.0,65.0)(65.0,73.0) 
\Text(30.0,57.0)[r]{$\gamma$}
\DashLine(31.0,65.0)(31.0,49.0){3.0} 
\Text(13.0,49.0)[r]{$\bar{e}$}
\ArrowLine(31.0,49.0)(14.0,49.0) 
\Text(39.0,53.0)[b]{$e$}
\ArrowLine(48.0,49.0)(31.0,49.0) 
\Text(66.0,57.0)[l]{$\bar{e}$}
\ArrowLine(65.0,57.0)(48.0,49.0) 
\Text(47.0,41.0)[r]{$Z$}
\DashLine(48.0,49.0)(48.0,33.0){3.0} 
\Text(66.0,41.0)[l]{$\mu$}
\ArrowLine(48.0,33.0)(65.0,41.0) 
\Text(66.0,25.0)[l]{$\bar{\mu}$}
\ArrowLine(65.0,25.0)(48.0,33.0) 
\Text(39,0)[b] {diagr.16}
\end{picture} \ 
{} \qquad\allowbreak
\begin{picture}(79,81)(0,0)
\Text(13.0,73.0)[r]{$e$}
\ArrowLine(14.0,73.0)(31.0,65.0) 
\Text(13.0,57.0)[r]{$\bar{e}$}
\ArrowLine(31.0,65.0)(14.0,57.0) 
\Text(39.0,66.0)[b]{$\gamma$}
\DashLine(31.0,65.0)(48.0,65.0){3.0} 
\Text(66.0,73.0)[l]{$\bar{\mu}$}
\ArrowLine(65.0,73.0)(48.0,65.0) 
\Text(44.0,57.0)[r]{$\mu$}
\ArrowLine(48.0,65.0)(48.0,49.0) 
\Text(66.0,57.0)[l]{$\mu$}
\ArrowLine(48.0,49.0)(65.0,57.0) 
\Text(47.0,41.0)[r]{$\gamma$}
\DashLine(48.0,49.0)(48.0,33.0){3.0} 
\Text(66.0,41.0)[l]{$e$}
\ArrowLine(48.0,33.0)(65.0,41.0) 
\Text(66.0,25.0)[l]{$\bar{e}$}
\ArrowLine(65.0,25.0)(48.0,33.0) 
\Text(39,0)[b] {diagr.17}
\end{picture} \ 
{} \qquad\allowbreak
\begin{picture}(79,81)(0,0)
\Text(13.0,73.0)[r]{$e$}
\ArrowLine(14.0,73.0)(48.0,73.0) 
\Text(66.0,73.0)[l]{$e$}
\ArrowLine(48.0,73.0)(65.0,73.0) 
\Text(47.0,65.0)[r]{$\gamma$}
\DashLine(48.0,73.0)(48.0,57.0){3.0} 
\Text(66.0,57.0)[l]{$\bar{\mu}$}
\ArrowLine(65.0,57.0)(48.0,57.0) 
\Text(44.0,49.0)[r]{$\mu$}
\ArrowLine(48.0,57.0)(48.0,41.0) 
\Text(66.0,41.0)[l]{$\mu$}
\ArrowLine(48.0,41.0)(65.0,41.0) 
\Text(47.0,33.0)[r]{$\gamma$}
\DashLine(48.0,41.0)(48.0,25.0){3.0} 
\Text(13.0,25.0)[r]{$\bar{e}$}
\ArrowLine(48.0,25.0)(14.0,25.0) 
\Text(66.0,25.0)[l]{$\bar{e}$}
\ArrowLine(65.0,25.0)(48.0,25.0) 
\Text(39,0)[b] {diagr.18}
\end{picture} \ 
{} \qquad\allowbreak
\begin{picture}(79,81)(0,0)
\Text(13.0,73.0)[r]{$e$}
\ArrowLine(14.0,73.0)(31.0,65.0) 
\Text(13.0,57.0)[r]{$\bar{e}$}
\ArrowLine(31.0,65.0)(14.0,57.0) 
\Text(39.0,66.0)[b]{$\gamma$}
\DashLine(31.0,65.0)(48.0,65.0){3.0} 
\Text(66.0,73.0)[l]{$\bar{\mu}$}
\ArrowLine(65.0,73.0)(48.0,65.0) 
\Text(44.0,57.0)[r]{$\mu$}
\ArrowLine(48.0,65.0)(48.0,49.0) 
\Text(66.0,57.0)[l]{$\mu$}
\ArrowLine(48.0,49.0)(65.0,57.0) 
\Text(47.0,41.0)[r]{$Z$}
\DashLine(48.0,49.0)(48.0,33.0){3.0} 
\Text(66.0,41.0)[l]{$e$}
\ArrowLine(48.0,33.0)(65.0,41.0) 
\Text(66.0,25.0)[l]{$\bar{e}$}
\ArrowLine(65.0,25.0)(48.0,33.0) 
\Text(39,0)[b] {diagr.19}
\end{picture} \ 
{} \qquad\allowbreak
\begin{picture}(79,81)(0,0)
\Text(13.0,73.0)[r]{$e$}
\ArrowLine(14.0,73.0)(48.0,73.0) 
\Text(66.0,73.0)[l]{$e$}
\ArrowLine(48.0,73.0)(65.0,73.0) 
\Text(47.0,65.0)[r]{$\gamma$}
\DashLine(48.0,73.0)(48.0,57.0){3.0} 
\Text(66.0,57.0)[l]{$\bar{\mu}$}
\ArrowLine(65.0,57.0)(48.0,57.0) 
\Text(44.0,49.0)[r]{$\mu$}
\ArrowLine(48.0,57.0)(48.0,41.0) 
\Text(66.0,41.0)[l]{$\mu$}
\ArrowLine(48.0,41.0)(65.0,41.0) 
\Text(47.0,33.0)[r]{$Z$}
\DashLine(48.0,41.0)(48.0,25.0){3.0} 
\Text(13.0,25.0)[r]{$\bar{e}$}
\ArrowLine(48.0,25.0)(14.0,25.0) 
\Text(66.0,25.0)[l]{$\bar{e}$}
\ArrowLine(65.0,25.0)(48.0,25.0) 
\Text(39,0)[b] {diagr.20}
\end{picture} \ 
{} \qquad\allowbreak
\begin{picture}(79,81)(0,0)
\Text(13.0,73.0)[r]{$e$}
\ArrowLine(14.0,73.0)(31.0,65.0) 
\Text(13.0,57.0)[r]{$\bar{e}$}
\ArrowLine(31.0,65.0)(14.0,57.0) 
\Text(39.0,66.0)[b]{$\gamma$}
\DashLine(31.0,65.0)(48.0,65.0){3.0} 
\Text(66.0,73.0)[l]{$\mu$}
\ArrowLine(48.0,65.0)(65.0,73.0) 
\Text(44.0,57.0)[r]{$\mu$}
\ArrowLine(48.0,49.0)(48.0,65.0) 
\Text(66.0,57.0)[l]{$\bar{\mu}$}
\ArrowLine(65.0,57.0)(48.0,49.0) 
\Text(47.0,41.0)[r]{$\gamma$}
\DashLine(48.0,49.0)(48.0,33.0){3.0} 
\Text(66.0,41.0)[l]{$e$}
\ArrowLine(48.0,33.0)(65.0,41.0) 
\Text(66.0,25.0)[l]{$\bar{e}$}
\ArrowLine(65.0,25.0)(48.0,33.0) 
\Text(39,0)[b] {diagr.21}
\end{picture} \ 
{} \qquad\allowbreak
\begin{picture}(79,81)(0,0)
\Text(13.0,73.0)[r]{$e$}
\ArrowLine(14.0,73.0)(48.0,73.0) 
\Text(66.0,73.0)[l]{$e$}
\ArrowLine(48.0,73.0)(65.0,73.0) 
\Text(47.0,65.0)[r]{$\gamma$}
\DashLine(48.0,73.0)(48.0,57.0){3.0} 
\Text(66.0,57.0)[l]{$\mu$}
\ArrowLine(48.0,57.0)(65.0,57.0) 
\Text(44.0,49.0)[r]{$\mu$}
\ArrowLine(48.0,41.0)(48.0,57.0) 
\Text(66.0,41.0)[l]{$\bar{\mu}$}
\ArrowLine(65.0,41.0)(48.0,41.0) 
\Text(47.0,33.0)[r]{$\gamma$}
\DashLine(48.0,41.0)(48.0,25.0){3.0} 
\Text(13.0,25.0)[r]{$\bar{e}$}
\ArrowLine(48.0,25.0)(14.0,25.0) 
\Text(66.0,25.0)[l]{$\bar{e}$}
\ArrowLine(65.0,25.0)(48.0,25.0) 
\Text(39,0)[b] {diagr.22}
\end{picture} \ 
{} \qquad\allowbreak
\begin{picture}(79,81)(0,0)
\Text(13.0,73.0)[r]{$e$}
\ArrowLine(14.0,73.0)(31.0,65.0) 
\Text(13.0,57.0)[r]{$\bar{e}$}
\ArrowLine(31.0,65.0)(14.0,57.0) 
\Text(39.0,66.0)[b]{$\gamma$}
\DashLine(31.0,65.0)(48.0,65.0){3.0} 
\Text(66.0,73.0)[l]{$\mu$}
\ArrowLine(48.0,65.0)(65.0,73.0) 
\Text(44.0,57.0)[r]{$\mu$}
\ArrowLine(48.0,49.0)(48.0,65.0) 
\Text(66.0,57.0)[l]{$\bar{\mu}$}
\ArrowLine(65.0,57.0)(48.0,49.0) 
\Text(47.0,41.0)[r]{$Z$}
\DashLine(48.0,49.0)(48.0,33.0){3.0} 
\Text(66.0,41.0)[l]{$e$}
\ArrowLine(48.0,33.0)(65.0,41.0) 
\Text(66.0,25.0)[l]{$\bar{e}$}
\ArrowLine(65.0,25.0)(48.0,33.0) 
\Text(39,0)[b] {diagr.23}
\end{picture} \ 
{} \qquad\allowbreak
\begin{picture}(79,81)(0,0)
\Text(13.0,73.0)[r]{$e$}
\ArrowLine(14.0,73.0)(48.0,73.0) 
\Text(66.0,73.0)[l]{$e$}
\ArrowLine(48.0,73.0)(65.0,73.0) 
\Text(47.0,65.0)[r]{$\gamma$}
\DashLine(48.0,73.0)(48.0,57.0){3.0} 
\Text(66.0,57.0)[l]{$\mu$}
\ArrowLine(48.0,57.0)(65.0,57.0) 
\Text(44.0,49.0)[r]{$\mu$}
\ArrowLine(48.0,41.0)(48.0,57.0) 
\Text(66.0,41.0)[l]{$\bar{\mu}$}
\ArrowLine(65.0,41.0)(48.0,41.0) 
\Text(47.0,33.0)[r]{$Z$}
\DashLine(48.0,41.0)(48.0,25.0){3.0} 
\Text(13.0,25.0)[r]{$\bar{e}$}
\ArrowLine(48.0,25.0)(14.0,25.0) 
\Text(66.0,25.0)[l]{$\bar{e}$}
\ArrowLine(65.0,25.0)(48.0,25.0) 
\Text(39,0)[b] {diagr.24}
\end{picture} \ 
{} \qquad\allowbreak
\begin{picture}(79,81)(0,0)
\Text(13.0,73.0)[r]{$e$}
\ArrowLine(14.0,73.0)(31.0,65.0) 
\Text(13.0,57.0)[r]{$\bar{e}$}
\ArrowLine(31.0,65.0)(14.0,57.0) 
\Text(39.0,66.0)[b]{$Z$}
\DashLine(31.0,65.0)(48.0,65.0){3.0} 
\Text(66.0,73.0)[l]{$\bar{e}$}
\ArrowLine(65.0,73.0)(48.0,65.0) 
\Text(44.0,57.0)[r]{$e$}
\ArrowLine(48.0,65.0)(48.0,49.0) 
\Text(66.0,57.0)[l]{$e$}
\ArrowLine(48.0,49.0)(65.0,57.0) 
\Text(47.0,41.0)[r]{$\gamma$}
\DashLine(48.0,49.0)(48.0,33.0){3.0} 
\Text(66.0,41.0)[l]{$\mu$}
\ArrowLine(48.0,33.0)(65.0,41.0) 
\Text(66.0,25.0)[l]{$\bar{\mu}$}
\ArrowLine(65.0,25.0)(48.0,33.0) 
\Text(39,0)[b] {diagr.25}
\end{picture} \ 
{} \qquad\allowbreak
\begin{picture}(79,81)(0,0)
\Text(13.0,65.0)[r]{$e$}
\ArrowLine(14.0,65.0)(31.0,65.0) 
\Line(31.0,65.0)(48.0,65.0) 
\Text(66.0,73.0)[l]{$e$}
\ArrowLine(48.0,65.0)(65.0,73.0) 
\Text(30.0,57.0)[r]{$Z$}
\DashLine(31.0,65.0)(31.0,49.0){3.0} 
\Line(31.0,49.0)(48.0,49.0) 
\Text(66.0,57.0)[l]{$\bar{e}$}
\ArrowLine(65.0,57.0)(48.0,49.0) 
\Text(27.0,41.0)[r]{$e$}
\ArrowLine(31.0,49.0)(31.0,33.0) 
\Text(13.0,33.0)[r]{$\bar{e}$}
\ArrowLine(31.0,33.0)(14.0,33.0) 
\Text(39.0,34.0)[b]{$\gamma$}
\DashLine(31.0,33.0)(48.0,33.0){3.0} 
\Text(66.0,41.0)[l]{$\mu$}
\ArrowLine(48.0,33.0)(65.0,41.0) 
\Text(66.0,25.0)[l]{$\bar{\mu}$}
\ArrowLine(65.0,25.0)(48.0,33.0) 
\Text(39,0)[b] {diagr.26}
\end{picture} \ 
{} \qquad\allowbreak
\begin{picture}(79,81)(0,0)
\Text(13.0,73.0)[r]{$e$}
\ArrowLine(14.0,73.0)(31.0,65.0) 
\Text(13.0,57.0)[r]{$\bar{e}$}
\ArrowLine(31.0,65.0)(14.0,57.0) 
\Text(39.0,66.0)[b]{$Z$}
\DashLine(31.0,65.0)(48.0,65.0){3.0} 
\Text(66.0,73.0)[l]{$\bar{e}$}
\ArrowLine(65.0,73.0)(48.0,65.0) 
\Text(44.0,57.0)[r]{$e$}
\ArrowLine(48.0,65.0)(48.0,49.0) 
\Text(66.0,57.0)[l]{$e$}
\ArrowLine(48.0,49.0)(65.0,57.0) 
\Text(47.0,41.0)[r]{$Z$}
\DashLine(48.0,49.0)(48.0,33.0){3.0} 
\Text(66.0,41.0)[l]{$\mu$}
\ArrowLine(48.0,33.0)(65.0,41.0) 
\Text(66.0,25.0)[l]{$\bar{\mu}$}
\ArrowLine(65.0,25.0)(48.0,33.0) 
\Text(39,0)[b] {diagr.27}
\end{picture} \ 
{} \qquad\allowbreak
\begin{picture}(79,81)(0,0)
\Text(13.0,65.0)[r]{$e$}
\ArrowLine(14.0,65.0)(31.0,65.0) 
\Line(31.0,65.0)(48.0,65.0) 
\Text(66.0,73.0)[l]{$e$}
\ArrowLine(48.0,65.0)(65.0,73.0) 
\Text(30.0,57.0)[r]{$Z$}
\DashLine(31.0,65.0)(31.0,49.0){3.0} 
\Line(31.0,49.0)(48.0,49.0) 
\Text(66.0,57.0)[l]{$\bar{e}$}
\ArrowLine(65.0,57.0)(48.0,49.0) 
\Text(27.0,41.0)[r]{$e$}
\ArrowLine(31.0,49.0)(31.0,33.0) 
\Text(13.0,33.0)[r]{$\bar{e}$}
\ArrowLine(31.0,33.0)(14.0,33.0) 
\Text(39.0,34.0)[b]{$Z$}
\DashLine(31.0,33.0)(48.0,33.0){3.0} 
\Text(66.0,41.0)[l]{$\mu$}
\ArrowLine(48.0,33.0)(65.0,41.0) 
\Text(66.0,25.0)[l]{$\bar{\mu}$}
\ArrowLine(65.0,25.0)(48.0,33.0) 
\Text(39,0)[b] {diagr.28}
\end{picture} \ 
{} \qquad\allowbreak
\begin{picture}(79,81)(0,0)
\Text(13.0,73.0)[r]{$e$}
\ArrowLine(14.0,73.0)(31.0,65.0) 
\Text(13.0,57.0)[r]{$\bar{e}$}
\ArrowLine(31.0,65.0)(14.0,57.0) 
\Text(39.0,66.0)[b]{$Z$}
\DashLine(31.0,65.0)(48.0,65.0){3.0} 
\Text(66.0,73.0)[l]{$e$}
\ArrowLine(48.0,65.0)(65.0,73.0) 
\Text(44.0,57.0)[r]{$e$}
\ArrowLine(48.0,49.0)(48.0,65.0) 
\Text(66.0,57.0)[l]{$\bar{e}$}
\ArrowLine(65.0,57.0)(48.0,49.0) 
\Text(47.0,41.0)[r]{$\gamma$}
\DashLine(48.0,49.0)(48.0,33.0){3.0} 
\Text(66.0,41.0)[l]{$\mu$}
\ArrowLine(48.0,33.0)(65.0,41.0) 
\Text(66.0,25.0)[l]{$\bar{\mu}$}
\ArrowLine(65.0,25.0)(48.0,33.0) 
\Text(39,0)[b] {diagr.29}
\end{picture} \ 
{} \qquad\allowbreak
\begin{picture}(79,81)(0,0)
\Text(13.0,65.0)[r]{$e$}
\ArrowLine(14.0,65.0)(31.0,65.0) 
\Line(31.0,65.0)(48.0,65.0) 
\Text(66.0,73.0)[l]{$e$}
\ArrowLine(48.0,65.0)(65.0,73.0) 
\Text(30.0,57.0)[r]{$Z$}
\DashLine(31.0,65.0)(31.0,49.0){3.0} 
\Text(13.0,49.0)[r]{$\bar{e}$}
\ArrowLine(31.0,49.0)(14.0,49.0) 
\Text(39.0,53.0)[b]{$e$}
\ArrowLine(48.0,49.0)(31.0,49.0) 
\Text(66.0,57.0)[l]{$\bar{e}$}
\ArrowLine(65.0,57.0)(48.0,49.0) 
\Text(47.0,41.0)[r]{$\gamma$}
\DashLine(48.0,49.0)(48.0,33.0){3.0} 
\Text(66.0,41.0)[l]{$\mu$}
\ArrowLine(48.0,33.0)(65.0,41.0) 
\Text(66.0,25.0)[l]{$\bar{\mu}$}
\ArrowLine(65.0,25.0)(48.0,33.0) 
\Text(39,0)[b] {diagr.30}
\end{picture} \ 
{} \qquad\allowbreak
\begin{picture}(79,81)(0,0)
\Text(13.0,73.0)[r]{$e$}
\ArrowLine(14.0,73.0)(31.0,65.0) 
\Text(13.0,57.0)[r]{$\bar{e}$}
\ArrowLine(31.0,65.0)(14.0,57.0) 
\Text(39.0,66.0)[b]{$Z$}
\DashLine(31.0,65.0)(48.0,65.0){3.0} 
\Text(66.0,73.0)[l]{$e$}
\ArrowLine(48.0,65.0)(65.0,73.0) 
\Text(44.0,57.0)[r]{$e$}
\ArrowLine(48.0,49.0)(48.0,65.0) 
\Text(66.0,57.0)[l]{$\bar{e}$}
\ArrowLine(65.0,57.0)(48.0,49.0) 
\Text(47.0,41.0)[r]{$Z$}
\DashLine(48.0,49.0)(48.0,33.0){3.0} 
\Text(66.0,41.0)[l]{$\mu$}
\ArrowLine(48.0,33.0)(65.0,41.0) 
\Text(66.0,25.0)[l]{$\bar{\mu}$}
\ArrowLine(65.0,25.0)(48.0,33.0) 
\Text(39,0)[b] {diagr.31}
\end{picture} \ 
{} \qquad\allowbreak
\begin{picture}(79,81)(0,0)
\Text(13.0,65.0)[r]{$e$}
\ArrowLine(14.0,65.0)(31.0,65.0) 
\Line(31.0,65.0)(48.0,65.0) 
\Text(66.0,73.0)[l]{$e$}
\ArrowLine(48.0,65.0)(65.0,73.0) 
\Text(30.0,57.0)[r]{$Z$}
\DashLine(31.0,65.0)(31.0,49.0){3.0} 
\Text(13.0,49.0)[r]{$\bar{e}$}
\ArrowLine(31.0,49.0)(14.0,49.0) 
\Text(39.0,53.0)[b]{$e$}
\ArrowLine(48.0,49.0)(31.0,49.0) 
\Text(66.0,57.0)[l]{$\bar{e}$}
\ArrowLine(65.0,57.0)(48.0,49.0) 
\Text(47.0,41.0)[r]{$Z$}
\DashLine(48.0,49.0)(48.0,33.0){3.0} 
\Text(66.0,41.0)[l]{$\mu$}
\ArrowLine(48.0,33.0)(65.0,41.0) 
\Text(66.0,25.0)[l]{$\bar{\mu}$}
\ArrowLine(65.0,25.0)(48.0,33.0) 
\Text(39,0)[b] {diagr.32}
\end{picture} \ 
{} \qquad\allowbreak
\begin{picture}(79,81)(0,0)
\Text(13.0,73.0)[r]{$e$}
\ArrowLine(14.0,73.0)(31.0,65.0) 
\Text(13.0,57.0)[r]{$\bar{e}$}
\ArrowLine(31.0,65.0)(14.0,57.0) 
\Text(39.0,66.0)[b]{$Z$}
\DashLine(31.0,65.0)(48.0,65.0){3.0} 
\Text(66.0,73.0)[l]{$\bar{\mu}$}
\ArrowLine(65.0,73.0)(48.0,65.0) 
\Text(44.0,57.0)[r]{$\mu$}
\ArrowLine(48.0,65.0)(48.0,49.0) 
\Text(66.0,57.0)[l]{$\mu$}
\ArrowLine(48.0,49.0)(65.0,57.0) 
\Text(47.0,41.0)[r]{$\gamma$}
\DashLine(48.0,49.0)(48.0,33.0){3.0} 
\Text(66.0,41.0)[l]{$e$}
\ArrowLine(48.0,33.0)(65.0,41.0) 
\Text(66.0,25.0)[l]{$\bar{e}$}
\ArrowLine(65.0,25.0)(48.0,33.0) 
\Text(39,0)[b] {diagr.33}
\end{picture} \ 
{} \qquad\allowbreak
\begin{picture}(79,81)(0,0)
\Text(13.0,73.0)[r]{$e$}
\ArrowLine(14.0,73.0)(48.0,73.0) 
\Text(66.0,73.0)[l]{$e$}
\ArrowLine(48.0,73.0)(65.0,73.0) 
\Text(47.0,65.0)[r]{$Z$}
\DashLine(48.0,73.0)(48.0,57.0){3.0} 
\Text(66.0,57.0)[l]{$\bar{\mu}$}
\ArrowLine(65.0,57.0)(48.0,57.0) 
\Text(44.0,49.0)[r]{$\mu$}
\ArrowLine(48.0,57.0)(48.0,41.0) 
\Text(66.0,41.0)[l]{$\mu$}
\ArrowLine(48.0,41.0)(65.0,41.0) 
\Text(47.0,33.0)[r]{$\gamma$}
\DashLine(48.0,41.0)(48.0,25.0){3.0} 
\Text(13.0,25.0)[r]{$\bar{e}$}
\ArrowLine(48.0,25.0)(14.0,25.0) 
\Text(66.0,25.0)[l]{$\bar{e}$}
\ArrowLine(65.0,25.0)(48.0,25.0) 
\Text(39,0)[b] {diagr.34}
\end{picture} \ 
{} \qquad\allowbreak
\begin{picture}(79,81)(0,0)
\Text(13.0,73.0)[r]{$e$}
\ArrowLine(14.0,73.0)(31.0,65.0) 
\Text(13.0,57.0)[r]{$\bar{e}$}
\ArrowLine(31.0,65.0)(14.0,57.0) 
\Text(39.0,66.0)[b]{$Z$}
\DashLine(31.0,65.0)(48.0,65.0){3.0} 
\Text(66.0,73.0)[l]{$\bar{\mu}$}
\ArrowLine(65.0,73.0)(48.0,65.0) 
\Text(44.0,57.0)[r]{$\mu$}
\ArrowLine(48.0,65.0)(48.0,49.0) 
\Text(66.0,57.0)[l]{$\mu$}
\ArrowLine(48.0,49.0)(65.0,57.0) 
\Text(47.0,41.0)[r]{$Z$}
\DashLine(48.0,49.0)(48.0,33.0){3.0} 
\Text(66.0,41.0)[l]{$e$}
\ArrowLine(48.0,33.0)(65.0,41.0) 
\Text(66.0,25.0)[l]{$\bar{e}$}
\ArrowLine(65.0,25.0)(48.0,33.0) 
\Text(39,0)[b] {diagr.35}
\end{picture} \ 
{} \qquad\allowbreak
\begin{picture}(79,81)(0,0)
\Text(13.0,73.0)[r]{$e$}
\ArrowLine(14.0,73.0)(48.0,73.0) 
\Text(66.0,73.0)[l]{$e$}
\ArrowLine(48.0,73.0)(65.0,73.0) 
\Text(47.0,65.0)[r]{$Z$}
\DashLine(48.0,73.0)(48.0,57.0){3.0} 
\Text(66.0,57.0)[l]{$\bar{\mu}$}
\ArrowLine(65.0,57.0)(48.0,57.0) 
\Text(44.0,49.0)[r]{$\mu$}
\ArrowLine(48.0,57.0)(48.0,41.0) 
\Text(66.0,41.0)[l]{$\mu$}
\ArrowLine(48.0,41.0)(65.0,41.0) 
\Text(47.0,33.0)[r]{$Z$}
\DashLine(48.0,41.0)(48.0,25.0){3.0} 
\Text(13.0,25.0)[r]{$\bar{e}$}
\ArrowLine(48.0,25.0)(14.0,25.0) 
\Text(66.0,25.0)[l]{$\bar{e}$}
\ArrowLine(65.0,25.0)(48.0,25.0) 
\Text(39,0)[b] {diagr.36}
\end{picture} \ 
{} \qquad\allowbreak
\begin{picture}(79,81)(0,0)
\Text(13.0,73.0)[r]{$e$}
\ArrowLine(14.0,73.0)(31.0,65.0) 
\Text(13.0,57.0)[r]{$\bar{e}$}
\ArrowLine(31.0,65.0)(14.0,57.0) 
\Text(39.0,66.0)[b]{$Z$}
\DashLine(31.0,65.0)(48.0,65.0){3.0} 
\Text(66.0,73.0)[l]{$\mu$}
\ArrowLine(48.0,65.0)(65.0,73.0) 
\Text(44.0,57.0)[r]{$\mu$}
\ArrowLine(48.0,49.0)(48.0,65.0) 
\Text(66.0,57.0)[l]{$\bar{\mu}$}
\ArrowLine(65.0,57.0)(48.0,49.0) 
\Text(47.0,41.0)[r]{$\gamma$}
\DashLine(48.0,49.0)(48.0,33.0){3.0} 
\Text(66.0,41.0)[l]{$e$}
\ArrowLine(48.0,33.0)(65.0,41.0) 
\Text(66.0,25.0)[l]{$\bar{e}$}
\ArrowLine(65.0,25.0)(48.0,33.0) 
\Text(39,0)[b] {diagr.37}
\end{picture} \ 
{} \qquad\allowbreak
\begin{picture}(79,81)(0,0)
\Text(13.0,73.0)[r]{$e$}
\ArrowLine(14.0,73.0)(48.0,73.0) 
\Text(66.0,73.0)[l]{$e$}
\ArrowLine(48.0,73.0)(65.0,73.0) 
\Text(47.0,65.0)[r]{$Z$}
\DashLine(48.0,73.0)(48.0,57.0){3.0} 
\Text(66.0,57.0)[l]{$\mu$}
\ArrowLine(48.0,57.0)(65.0,57.0) 
\Text(44.0,49.0)[r]{$\mu$}
\ArrowLine(48.0,41.0)(48.0,57.0) 
\Text(66.0,41.0)[l]{$\bar{\mu}$}
\ArrowLine(65.0,41.0)(48.0,41.0) 
\Text(47.0,33.0)[r]{$\gamma$}
\DashLine(48.0,41.0)(48.0,25.0){3.0} 
\Text(13.0,25.0)[r]{$\bar{e}$}
\ArrowLine(48.0,25.0)(14.0,25.0) 
\Text(66.0,25.0)[l]{$\bar{e}$}
\ArrowLine(65.0,25.0)(48.0,25.0) 
\Text(39,0)[b] {diagr.38}
\end{picture} \ 
{} \qquad\allowbreak
\begin{picture}(79,81)(0,0)
\Text(13.0,73.0)[r]{$e$}
\ArrowLine(14.0,73.0)(31.0,65.0) 
\Text(13.0,57.0)[r]{$\bar{e}$}
\ArrowLine(31.0,65.0)(14.0,57.0) 
\Text(39.0,66.0)[b]{$Z$}
\DashLine(31.0,65.0)(48.0,65.0){3.0} 
\Text(66.0,73.0)[l]{$\mu$}
\ArrowLine(48.0,65.0)(65.0,73.0) 
\Text(44.0,57.0)[r]{$\mu$}
\ArrowLine(48.0,49.0)(48.0,65.0) 
\Text(66.0,57.0)[l]{$\bar{\mu}$}
\ArrowLine(65.0,57.0)(48.0,49.0) 
\Text(47.0,41.0)[r]{$Z$}
\DashLine(48.0,49.0)(48.0,33.0){3.0} 
\Text(66.0,41.0)[l]{$e$}
\ArrowLine(48.0,33.0)(65.0,41.0) 
\Text(66.0,25.0)[l]{$\bar{e}$}
\ArrowLine(65.0,25.0)(48.0,33.0) 
\Text(39,0)[b] {diagr.39}
\end{picture} \ 
{} \qquad\allowbreak
\begin{picture}(79,81)(0,0)
\Text(13.0,73.0)[r]{$e$}
\ArrowLine(14.0,73.0)(48.0,73.0) 
\Text(66.0,73.0)[l]{$e$}
\ArrowLine(48.0,73.0)(65.0,73.0) 
\Text(47.0,65.0)[r]{$Z$}
\DashLine(48.0,73.0)(48.0,57.0){3.0} 
\Text(66.0,57.0)[l]{$\mu$}
\ArrowLine(48.0,57.0)(65.0,57.0) 
\Text(44.0,49.0)[r]{$\mu$}
\ArrowLine(48.0,41.0)(48.0,57.0) 
\Text(66.0,41.0)[l]{$\bar{\mu}$}
\ArrowLine(65.0,41.0)(48.0,41.0) 
\Text(47.0,33.0)[r]{$Z$}
\DashLine(48.0,41.0)(48.0,25.0){3.0} 
\Text(13.0,25.0)[r]{$\bar{e}$}
\ArrowLine(48.0,25.0)(14.0,25.0) 
\Text(66.0,25.0)[l]{$\bar{e}$}
\ArrowLine(65.0,25.0)(48.0,25.0) 
\Text(39,0)[b] {diagr.40}
\end{picture} \ 
{} \qquad\allowbreak
\begin{picture}(79,81)(0,0)
\Text(13.0,57.0)[r]{$e$}
\ArrowLine(14.0,57.0)(31.0,49.0) 
\Text(13.0,41.0)[r]{$\bar{e}$}
\ArrowLine(31.0,49.0)(14.0,41.0) 
\Text(31.0,51.0)[lb]{$Z$}
\DashLine(31.0,49.0)(48.0,49.0){3.0} 
\Text(47.0,57.0)[r]{$H$}
\DashLine(48.0,49.0)(48.0,65.0){1.0}
\Text(66.0,73.0)[l]{$\mu$}
\ArrowLine(48.0,65.0)(65.0,73.0) 
\Text(66.0,57.0)[l]{$\bar{\mu}$}
\ArrowLine(65.0,57.0)(48.0,65.0) 
\Text(47.0,41.0)[r]{$Z$}
\DashLine(48.0,49.0)(48.0,33.0){3.0} 
\Text(66.0,41.0)[l]{$e$}
\ArrowLine(48.0,33.0)(65.0,41.0) 
\Text(66.0,25.0)[l]{$\bar{e}$}
\ArrowLine(65.0,25.0)(48.0,33.0) 
\Text(39,0)[b] {diagr.41}
\end{picture} \ 
{} \qquad\allowbreak
\begin{picture}(79,81)(0,0)
\Text(13.0,65.0)[r]{$e$}
\ArrowLine(14.0,65.0)(31.0,65.0) 
\Line(31.0,65.0)(48.0,65.0) 
\Text(66.0,73.0)[l]{$e$}
\ArrowLine(48.0,65.0)(65.0,73.0) 
\Text(30.0,57.0)[r]{$Z$}
\DashLine(31.0,65.0)(31.0,49.0){3.0} 
\Text(39.0,50.0)[b]{$H$}
\DashLine(31.0,49.0)(48.0,49.0){1.0}
\Text(66.0,57.0)[l]{$\mu$}
\ArrowLine(48.0,49.0)(65.0,57.0) 
\Text(66.0,41.0)[l]{$\bar{\mu}$}
\ArrowLine(65.0,41.0)(48.0,49.0) 
\Text(30.0,41.0)[r]{$Z$}
\DashLine(31.0,49.0)(31.0,33.0){3.0} 
\Text(13.0,33.0)[r]{$\bar{e}$}
\ArrowLine(31.0,33.0)(14.0,33.0) 
\Line(31.0,33.0)(48.0,33.0) 
\Text(66.0,25.0)[l]{$\bar{e}$}
\ArrowLine(65.0,25.0)(48.0,33.0) 
\Text(39,0)[b] {diagr.42}
\end{picture} \ 
{} \qquad\allowbreak
\begin{picture}(79,81)(0,0)
\Text(13.0,65.0)[r]{$e$}
\ArrowLine(14.0,65.0)(31.0,65.0) 
\Text(39.0,66.0)[b]{$Z$}
\DashLine(31.0,65.0)(48.0,65.0){3.0} 
\Text(66.0,73.0)[l]{$\mu$}
\ArrowLine(48.0,65.0)(65.0,73.0) 
\Text(66.0,57.0)[l]{$\bar{\mu}$}
\ArrowLine(65.0,57.0)(48.0,65.0) 
\Text(27.0,49.0)[r]{$e$}
\ArrowLine(31.0,65.0)(31.0,33.0) 
\Text(13.0,33.0)[r]{$\bar{e}$}
\ArrowLine(31.0,33.0)(14.0,33.0) 
\Text(39.0,34.0)[b]{$\gamma$}
\DashLine(31.0,33.0)(48.0,33.0){3.0} 
\Text(66.0,41.0)[l]{$e$}
\ArrowLine(48.0,33.0)(65.0,41.0) 
\Text(66.0,25.0)[l]{$\bar{e}$}
\ArrowLine(65.0,25.0)(48.0,33.0) 
\Text(39,0)[b] {diagr.43}
\end{picture} \ 
{} \qquad\allowbreak
\begin{picture}(79,81)(0,0)
\Text(13.0,65.0)[r]{$e$}
\ArrowLine(14.0,65.0)(31.0,65.0) 
\Text(39.0,66.0)[b]{$Z$}
\DashLine(31.0,65.0)(48.0,65.0){3.0} 
\Text(66.0,73.0)[l]{$\mu$}
\ArrowLine(48.0,65.0)(65.0,73.0) 
\Text(66.0,57.0)[l]{$\bar{\mu}$}
\ArrowLine(65.0,57.0)(48.0,65.0) 
\Text(27.0,57.0)[r]{$e$}
\ArrowLine(31.0,65.0)(31.0,49.0) 
\Line(31.0,49.0)(48.0,49.0) 
\Text(66.0,41.0)[l]{$e$}
\ArrowLine(48.0,49.0)(65.0,41.0) 
\Text(30.0,41.0)[r]{$\gamma$}
\DashLine(31.0,49.0)(31.0,33.0){3.0} 
\Text(13.0,33.0)[r]{$\bar{e}$}
\ArrowLine(31.0,33.0)(14.0,33.0) 
\Line(31.0,33.0)(48.0,33.0) 
\Text(66.0,25.0)[l]{$\bar{e}$}
\ArrowLine(65.0,25.0)(48.0,33.0) 
\Text(39,0)[b] {diagr.44}
\end{picture} \ 
{} \qquad\allowbreak
\begin{picture}(79,81)(0,0)
\Text(13.0,65.0)[r]{$e$}
\ArrowLine(14.0,65.0)(31.0,65.0) 
\Text(39.0,69.0)[b]{$e$}
\ArrowLine(31.0,65.0)(48.0,65.0) 
\Text(66.0,73.0)[l]{$e$}
\ArrowLine(48.0,65.0)(65.0,73.0) 
\Text(47.0,57.0)[r]{$\gamma$}
\DashLine(48.0,65.0)(48.0,49.0){3.0} 
\Text(66.0,57.0)[l]{$\mu$}
\ArrowLine(48.0,49.0)(65.0,57.0) 
\Text(66.0,41.0)[l]{$\bar{\mu}$}
\ArrowLine(65.0,41.0)(48.0,49.0) 
\Text(30.0,49.0)[r]{$Z$}
\DashLine(31.0,65.0)(31.0,33.0){3.0} 
\Text(13.0,33.0)[r]{$\bar{e}$}
\ArrowLine(31.0,33.0)(14.0,33.0) 
\Line(31.0,33.0)(48.0,33.0) 
\Text(66.0,25.0)[l]{$\bar{e}$}
\ArrowLine(65.0,25.0)(48.0,33.0) 
\Text(39,0)[b] {diagr.45}
\end{picture} \ 
{} \qquad\allowbreak
\begin{picture}(79,81)(0,0)
\Text(13.0,65.0)[r]{$e$}
\ArrowLine(14.0,65.0)(31.0,65.0) 
\Text(39.0,66.0)[b]{$Z$}
\DashLine(31.0,65.0)(48.0,65.0){3.0} 
\Text(66.0,73.0)[l]{$e$}
\ArrowLine(48.0,65.0)(65.0,73.0) 
\Text(66.0,57.0)[l]{$\bar{e}$}
\ArrowLine(65.0,57.0)(48.0,65.0) 
\Text(27.0,49.0)[r]{$e$}
\ArrowLine(31.0,65.0)(31.0,33.0) 
\Text(13.0,33.0)[r]{$\bar{e}$}
\ArrowLine(31.0,33.0)(14.0,33.0) 
\Text(39.0,34.0)[b]{$\gamma$}
\DashLine(31.0,33.0)(48.0,33.0){3.0} 
\Text(66.0,41.0)[l]{$\mu$}
\ArrowLine(48.0,33.0)(65.0,41.0) 
\Text(66.0,25.0)[l]{$\bar{\mu}$}
\ArrowLine(65.0,25.0)(48.0,33.0) 
\Text(39,0)[b] {diagr.46}
\end{picture} \ 
{} \qquad\allowbreak
\begin{picture}(79,81)(0,0)
\Text(13.0,65.0)[r]{$e$}
\ArrowLine(14.0,65.0)(31.0,65.0) 
\Text(39.0,66.0)[b]{$Z$}
\DashLine(31.0,65.0)(48.0,65.0){3.0} 
\Text(66.0,73.0)[l]{$\mu$}
\ArrowLine(48.0,65.0)(65.0,73.0) 
\Text(66.0,57.0)[l]{$\bar{\mu}$}
\ArrowLine(65.0,57.0)(48.0,65.0) 
\Text(27.0,49.0)[r]{$e$}
\ArrowLine(31.0,65.0)(31.0,33.0) 
\Text(13.0,33.0)[r]{$\bar{e}$}
\ArrowLine(31.0,33.0)(14.0,33.0) 
\Text(39.0,34.0)[b]{$Z$}
\DashLine(31.0,33.0)(48.0,33.0){3.0} 
\Text(66.0,41.0)[l]{$e$}
\ArrowLine(48.0,33.0)(65.0,41.0) 
\Text(66.0,25.0)[l]{$\bar{e}$}
\ArrowLine(65.0,25.0)(48.0,33.0) 
\Text(39,0)[b] {diagr.47}
\end{picture} \ 
{} \qquad\allowbreak
\begin{picture}(79,81)(0,0)
\Text(13.0,65.0)[r]{$e$}
\ArrowLine(14.0,65.0)(31.0,65.0) 
\Text(39.0,66.0)[b]{$Z$}
\DashLine(31.0,65.0)(48.0,65.0){3.0} 
\Text(66.0,73.0)[l]{$\mu$}
\ArrowLine(48.0,65.0)(65.0,73.0) 
\Text(66.0,57.0)[l]{$\bar{\mu}$}
\ArrowLine(65.0,57.0)(48.0,65.0) 
\Text(27.0,57.0)[r]{$e$}
\ArrowLine(31.0,65.0)(31.0,49.0) 
\Line(31.0,49.0)(48.0,49.0) 
\Text(66.0,41.0)[l]{$e$}
\ArrowLine(48.0,49.0)(65.0,41.0) 
\Text(30.0,41.0)[r]{$Z$}
\DashLine(31.0,49.0)(31.0,33.0){3.0} 
\Text(13.0,33.0)[r]{$\bar{e}$}
\ArrowLine(31.0,33.0)(14.0,33.0) 
\Line(31.0,33.0)(48.0,33.0) 
\Text(66.0,25.0)[l]{$\bar{e}$}
\ArrowLine(65.0,25.0)(48.0,33.0) 
\Text(39,0)[b] {diagr.48}
\end{picture} \ 
{} \qquad\allowbreak
\begin{picture}(79,81)(0,0)
\Text(13.0,65.0)[r]{$e$}
\ArrowLine(14.0,65.0)(31.0,65.0) 
\Text(39.0,69.0)[b]{$e$}
\ArrowLine(31.0,65.0)(48.0,65.0) 
\Text(66.0,73.0)[l]{$e$}
\ArrowLine(48.0,65.0)(65.0,73.0) 
\Text(47.0,57.0)[r]{$Z$}
\DashLine(48.0,65.0)(48.0,49.0){3.0} 
\Text(66.0,57.0)[l]{$\mu$}
\ArrowLine(48.0,49.0)(65.0,57.0) 
\Text(66.0,41.0)[l]{$\bar{\mu}$}
\ArrowLine(65.0,41.0)(48.0,49.0) 
\Text(30.0,49.0)[r]{$Z$}
\DashLine(31.0,65.0)(31.0,33.0){3.0} 
\Text(13.0,33.0)[r]{$\bar{e}$}
\ArrowLine(31.0,33.0)(14.0,33.0) 
\Line(31.0,33.0)(48.0,33.0) 
\Text(66.0,25.0)[l]{$\bar{e}$}
\ArrowLine(65.0,25.0)(48.0,33.0) 
\Text(39,0)[b] {diagr.49}
\end{picture} \ 
{} \qquad\allowbreak
\begin{picture}(79,81)(0,0)
\Text(13.0,65.0)[r]{$e$}
\ArrowLine(14.0,65.0)(31.0,65.0) 
\Text(39.0,66.0)[b]{$Z$}
\DashLine(31.0,65.0)(48.0,65.0){3.0} 
\Text(66.0,73.0)[l]{$e$}
\ArrowLine(48.0,65.0)(65.0,73.0) 
\Text(66.0,57.0)[l]{$\bar{e}$}
\ArrowLine(65.0,57.0)(48.0,65.0) 
\Text(27.0,49.0)[r]{$e$}
\ArrowLine(31.0,65.0)(31.0,33.0) 
\Text(13.0,33.0)[r]{$\bar{e}$}
\ArrowLine(31.0,33.0)(14.0,33.0) 
\Text(39.0,34.0)[b]{$Z$}
\DashLine(31.0,33.0)(48.0,33.0){3.0} 
\Text(66.0,41.0)[l]{$\mu$}
\ArrowLine(48.0,33.0)(65.0,41.0) 
\Text(66.0,25.0)[l]{$\bar{\mu}$}
\ArrowLine(65.0,25.0)(48.0,33.0) 
\Text(39,0)[b] {diagr.50}
\end{picture} \ 
\end{document}